\documentclass[12pt]{iopart}
\usepackage{iopams}
\usepackage{subfigure}
\usepackage{graphicx}
   
\newcommand{\be}{\begin{equation}}
\newcommand{\ee}{\end{equation}}
 
\begin{document}

\title{Interacting Particle Systems in Time-Dependent Geometries}

\author{A Ali$^1$, R Ball$^{1,3}$, S Grosskinsky$^{1,2}$ and E Somfai$^{1,3,4}$}
\address{$^1$ Centre for Complexity Science, University of Warwick,
Coventry CV4 7AL, United Kingdom\\
$^2$ Department of Mathematics, University of Warwick,
Coventry CV4 7AL, United Kingdom\\
$^3$ Department of Physics, University of Warwick,
Coventry CV4 7AL, United Kingdom\\
$^4$ Institute for Solid State Physics and Optics, Wigner Research Center for
Physics, Hungarian Academy of Sciences, P.O. Box 49, H-1525 Budapest, Hungary
}

\ead{S.W.Grosskinsky@warwick.ac.uk}
\begin{abstract}
Many complex structures and stochastic patterns emerge from simple kinetic rules and local interactions, and are governed by scale invariance properties in combination with effects of the global geometry. We consider systems that can be described effectively by space-time trajectories of interacting particles, such as domain boundaries in two-dimensional growth or river networks. We study trajectories embedded in 
time-dependent geometries, and the main focus is on uniformly expanding or decreasing domains 
for which we obtain an exact mapping to simple fixed domain systems while preserving the local scale invariance properties. This approach was recently introduced in [A.\ Ali et al., Phys.\ Rev.\ E \textbf{87}, 020102(R) (2013)] and here we provide a detailed discussion on its applicability for self-affince Markovian models, and how it can be adapted to self-affine models with memory or explicit time dependence. The mapping corresponds to a non-linear time transformation which convergences to a finite value for a large class of trajectories, enabling an exact analysis of asymptotic properties in expanding domains. We further provide a detailed discussion of different particle interactions and generalized geometries. All our findings are based on exact computations and are illustrated numerically for various examples, including L\'evy processes and fractional Brownian motion.

\end{abstract}

\maketitle

\section{Introduction}

Scale invariant random structures are common across various real systems and mathematical models, examples include diffusion limited aggregation (DLA) \cite{Witten1981,Stanley1990,Stanley1994}, domain boundaries of crystal growth \cite{Saito1995}, viscous fingering \cite{Saffman1958,Paterson1981,Thome1989b}, microbial growth \cite{Hallatschek2007,Korolev2010b} or the landscape of river networks \cite{Somfai1997}. In such systems scale invariance properties result from local stochastic growth or fluctuations due to localized inhomogeneities. 
These local effects lead to emerging patterns on larger length scales, in combination with global geometric properties and constraints of the system \cite{Barabasi1995,Grosskinsky2010,Ali2012}.
Understanding how these effects interplay to affect the macroscopic observables and pattern formation is of great interest 
in non-equilibrium statistical mechanics, and can lead to further understanding of fundamental processes, such as diffusion transport, diffusion controlled reactions and aggregation structure formation \cite{Barabasi1995,Bunde1993,Kopelman1988}.

In this paper we focus on phenomena that can be effectively described by space-time trajectories of interacting particles. This includes a variety of systems such as domain boundaries in two-dimensional competitive growth or river networks, crystal growth, liquid invasion in porous media, or epidemic spreading and microbial growth \cite{Fisher1989,Jacob2010,Murray2003}. 
Typical patterns observed can range from polyhedral, dendritic, fractal to compact structures 
\cite{Mandelbrot1982}. The overall geometry usually has a strong impact on the observed behaviour \cite{Paterson1981,Thome1989b,Lehe2012b,Lebovka1998} and often changes dynamically, which is the major interest of this paper and can play in fact a dominant role on the kinetics of fluctuating particles \cite{Grosskinsky2010,Eden1961,Lebovka1998}. We will focus mostly on simple contact interactions such as annihilation or coalescence, but will also discuss how to include more general interactions with their own length scale, such as branching. 

\begin{figure}
\begin{center}
\subfigure[]{\includegraphics[bb=105 276 489 565,clip,width=3.0in]{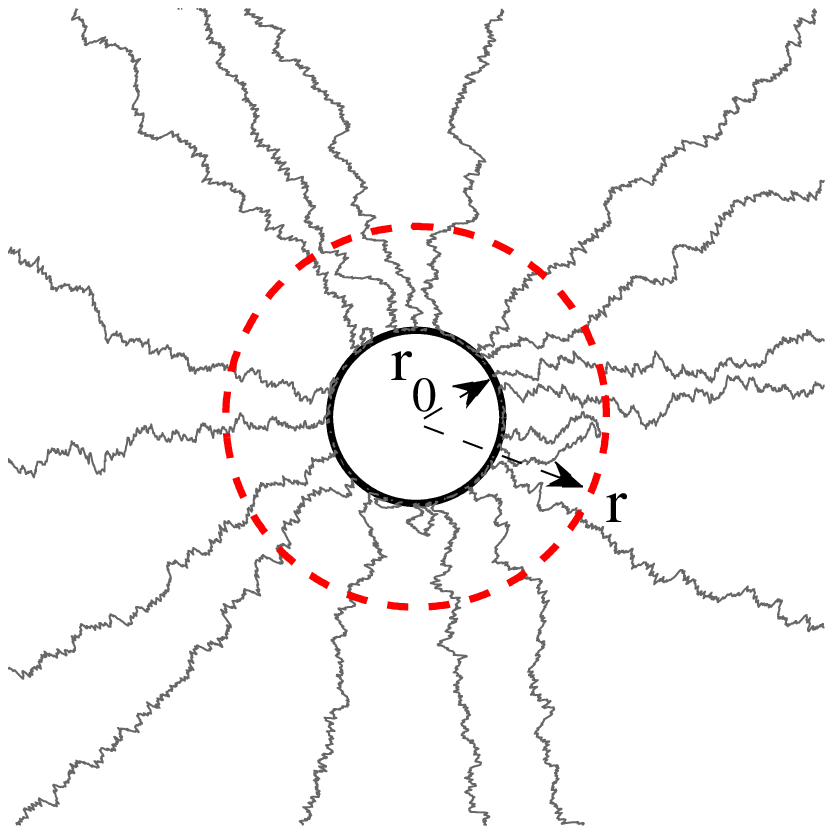}}
\subfigure[]{\includegraphics[bb=105 276 489 565,clip,width=3.0in]{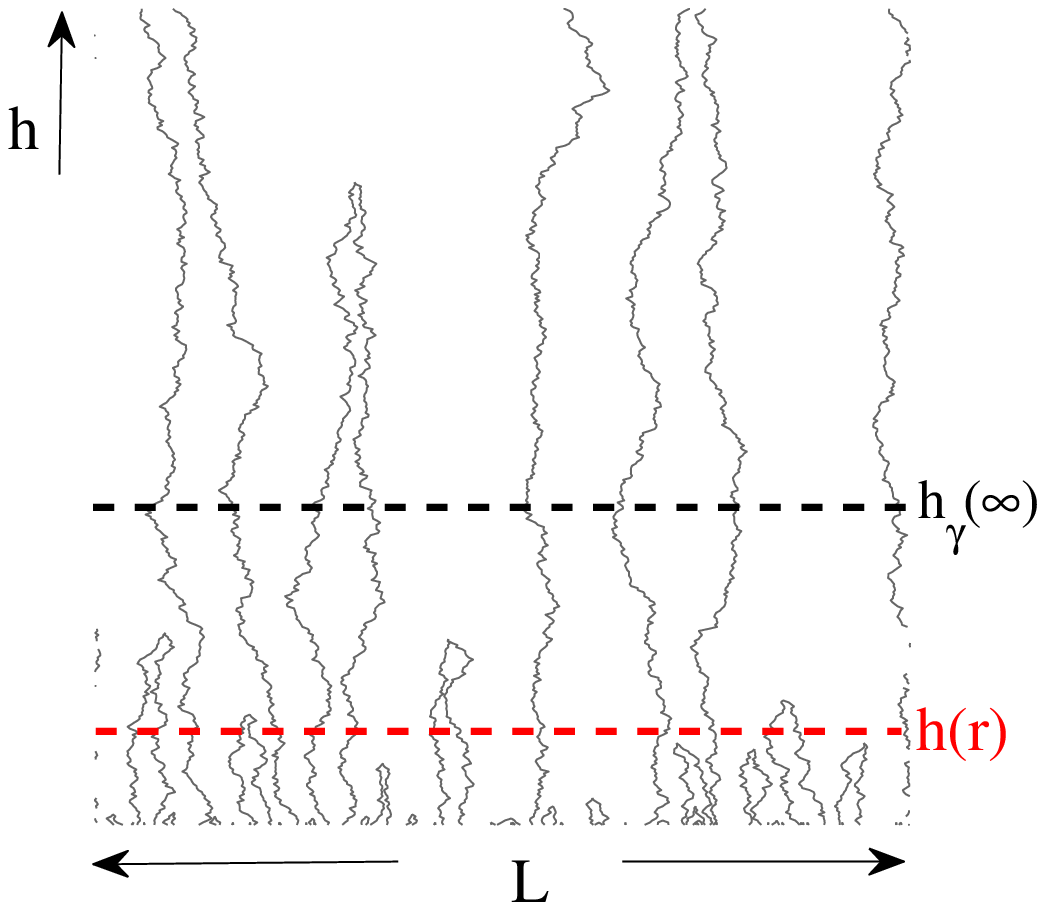}}
\end{center}
\caption{\label{bmmap}
Expanding radial growth structure (a) and the same structure on
a fixed domain with periodic boundary conditions (b), illustrated for annihilating Brownian motion ($\gamma =1/2$). The distribution of the rescaled structure
at radius $r$ is identical to the distribution of the fixed domain structure
at height $h(r)$ as given by the mapping (\ref{mapping}), indicated by a red
dashed line.
The mapping has a finite limit $h_\gamma (\infty )$ for $\gamma <1$, see also
Figure~\ref{figmapping}.
Parameters are $L=100$ with $r_0 = L/2\pi$, unit diffusion coefficient and initially $100$ arms.
}
\end{figure}

The main illustrative example will be trajectories in an expanding radial domain, which leads to different results as seen in fixed linear geometries; an example is shown in Figure~\ref{bmmap} for annihilating Brownian motions. 
We compare the behaviour on  radially expanding space (a) with a fixed space with periodic boundary conditions (b). The behaviour on the fixed domain is well understood for various interactions (see e.g. \cite{Lebovka1998,Munasinghe2006,Alemany1995,Sasaki2000,Avraham2001} and references therein) and we will use the local scale invariance properties of the model to map the behaviour from time dependent domains onto fixed domains. Focusing on uniformly expanding or decreasing domains, we derive a general mapping first published in \cite{Ali2013}, that universally applies to all particle trajectories with Markovian statistics, which we illustrate for Brownian motion and superdiffusive $\alpha$-stable L\'evy processes. We further show how to generalize these results to processes with memory, giving an explicit result for super- and subdiffusive fractional Brownian motion, and discuss Brownian motion with time-dependent diffusivity as an example of a self-similar but not locally scale invariant process.
These mappings can be interpreted as a non-linear time change of the rescaled processes in time-dependent geometry, which converges to a finite value if spatial expansion or decrease of the domain dominates the fluctuations of the trajectories. This leads to an exact prediction of asymptotic properties on time-dependent domains, which are mapped to a finite-time statistics of fixed domain systems. The latter often exhibit convergence to absorbing states as for example for annihilation or coagulation interactions, whereas on the time-dependent domain processes can show fluctuating limiting behaviour which depends crucially on the initial dynamics. For the fixed domain statistics such as the number of particles or the inter-particle distance are known as a function of time, and we are able to make predictions on not just the asymptotics but also on the dynamical behaviour. All our results follow from exact computations and detailed numerical simulations are performed mostly for illustration purposes.

The paper is organized as follows. In Section~\ref{secresults} we derive a general mapping based on the preservation of local scale invariance which depends only on the scaling exponent $\gamma$, 
and we give a detailed explanation on its properties. In Section~\ref{models} the mapping is applied to explicit models of radially expanding structures, consisting of point particle trajectories. We give a rigorous derivation of the mapping for the Markovian models of Brownian motion and $\alpha$-stable L\'evy processes, applying techniques of stochastic calculus. We also derive equivalent but different mappings for processes with memory, including fractional Brownian motion, using moment matching. In Section~\ref{generaldomain} we extend our theory to describe structures which reside on general time-dependent domains with isotropic evolution, which can also be higher dimensional. Lastly in Section~\ref{non-local}, we explain how to include non-local interactions in our approach, such as non-zero particle sizes or branching, which has its own characteristic time scale. 
These systems are more natural in the real world, see \cite{Saffman1958,Barabasi1995,Jacob2010,Murray2003,Pazsit2007} for a more general overview, and are thus an important adaption of our simple theory for point particles.

\section{Main results \label{secresults}}
We derive a mapping first given in \cite{Ali2013}, which is used to describe behaviour of expanding structures by mapping to structures in a fixed domain, as illustrated in Figure~\ref{bmmap}. Here for simplicity of presentation we focus on the radial geometry and compare it to the strip geometry with periodic boundary conditions. Extensions to more general geometries can be found in Section~\ref{generaldomain}.

\subsection{Mapping \label{secmapping}}

Consider an isotropic radial structure with initial radius $r_0$ as shown in
Figure~\ref{bmmap}(a), which is a particular example of radially annihilating Brownian
motion. We consider directed radial growth where each arm of the displacement along the perimeter of the
growing circle can be represented as a function of the radial distance $r$,
$$
  (Y_r, r\ge r_0)\quad \mbox{with} \quad Y_r \in [0,2\pi r )\ .
$$
Increments of this process
\begin{equation}\label{yin}
dY_r =Y_r \, dr/r +d\tilde Y_r
\end{equation}
are given by a contribution due to the stretching of space, and a second one due to the inherent fluctuations encoding
the local scale invariance of the arms. We use this notation for increments on a formal, heuristic level in this section, which is made mathematically precise in Section 3.
In the analogous fixed domain geometry Figure~\ref{bmmap}(b), we model a single arm of the same growth structure as a process
$$
(X_{h},h\ge 0) \quad \mbox{with} \quad X_{h} \in [0,L)\ , 
$$
for which the increments are simply given by fluctuations $dX_h$.
In order to connect the two domains we take $r_{0}=L/2\pi$ and have periodic boundary conditions at the edges. With matching the initial conditions $X_0 =Y_{r_0}$ this implies
\begin{equation} X_h =\frac{r_{0}}{r} Y_r \ , \end{equation}
in analogy to the usual polar coordinated transformation. The displacement of the rescaled radial arm has the same range as the fixed arm, and using (\ref{yin}) we get for the increments
\begin{equation} dX_h =\frac{r_{0}}{r} d\tilde Y_r \label{relation}\ .\end{equation}
In each geometry the arms share the same local scale invariance property, i.e. the increments due to fluctuations scale as
\begin{equation} dX_h \sim (dh)^{\gamma} \quad \mbox{and} \quad d\tilde Y_r \sim (dr)^{\gamma}, \label{locsca}\end{equation}
where $\gamma >0$ and proportionality constants in both cases are the same. 
Generic examples are self-similar processes where $(X_{bh}, h\ge0)$ is distributed as $(b^{\gamma}X_{h}, h\ge0)$ for all $b>0$, such as fractional Brownian motion \cite{Biagini2008}, or $\alpha$-stable L\'evy processes \cite{Chechkin2008}, which will be discussed in Section~\ref{models} in more detail. 

\begin{figure}[t]
\begin{center}
\includegraphics[width=4in]{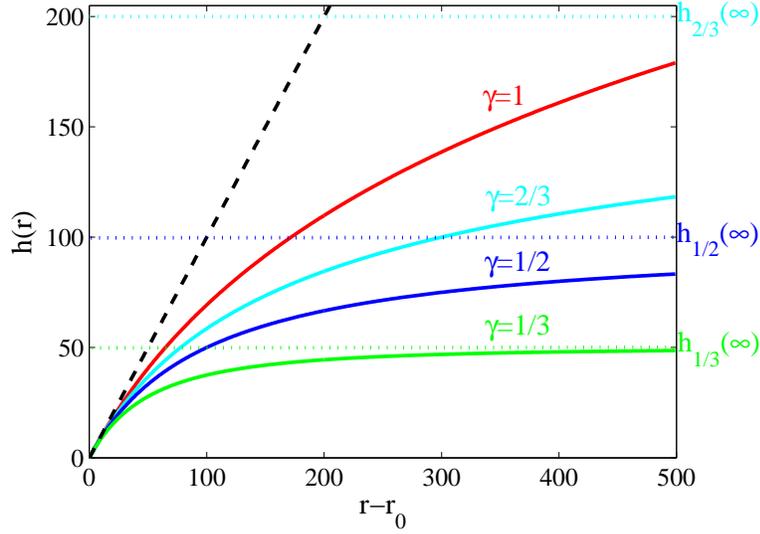}
\end{center}
\caption{\label{figmapping}
The mapping (\ref{mapping}), for several values of $\gamma$, with $r_{0}=100$. Initially the mapping $h(r)$ behaves as the identity $r-r_{0}$ (black dashed line) and converges to the limit $h_{\infty}(\gamma)=r_{0}\gamma/(1-\gamma)$ (color dotted) as $r\rightarrow \infty$. The value $h_{\infty}(\gamma)$ corresponds to the height $h$ in the fixed domain where the behaviour is equivalent to $r \rightarrow \infty$ in the radially growing structure. (See also Figure~\ref{bmref}).
}
\end{figure}

From (\ref{locsca}) and (\ref{relation}) we get heuristically
$$\frac{dh}{dr}= \Big(\frac{dx}{dy} \Big)^{1/\gamma} = \Big(\frac{r_{0}}{r}\Big)^{1/\gamma},$$
and therefore
\begin{equation}
h(r)=\int_{r_{0}}^{r}\Big(\frac{r_{0}}{s}\Big)^{1/\gamma} ds= \left\{\begin{array}{cl} \frac{\gamma }{1-\gamma }\, r_{0}\Big[1-\big(\frac{r_{0}}{r}\big)^{\frac{1-\gamma}{\gamma}} \Big] &,\ \gamma \ne 1  \\ r_{0}\log\big(\frac{r}{r_{0}}\big) &,\ \gamma=1 \end{array}\right. \label{mapping}.
\end{equation}
For a single arm matching the initial condition
$Y_{r_0} =X_0$ leads to the identical distribution 
\be\label{strongres}
\Big( \frac{r_0}{r}\, Y_r ,r\geq r_0\Big) \stackrel{\mathrm{dist.}}{=} \big(X_{h(r)} ,r\geq r_0 \big) ,
\ee
Crucially, the same holds for the entire growth structure
which are characterized as a collection of arms $\{ (Y_r ,r\geq r_0 )\}$ and $\{ (X_h ,h\geq 0)\}$:
\be
\Big\{\Big( \frac{r_0}{r}\, Y_r ,r\geq r_0\Big) \Big\}
\stackrel{\mathrm{dist.}}{=}\Big\{ \big(X_{h(r)} ,r\geq r_0 \big) \Big\} \ ,
\ee
provided that the arms interact locally in a scale independent way. Examples of such
interactions include coagulation, annihilation or exclusion, and extensions to non-local interactions are discussed in Section~\ref{non-local}. Figure~\ref{bmmap}
illustrates the correspondence given by the mapping for annihilating Brownian motion, where the red dashed line indicates where the marginal distributions of the two structures are equal. Note that in the above formulation equivalence holds for the full trajectories including time correlations. This will be confirmed rigorously in Section \ref{models} for Markovian models, and a weaker corrected version will be derived for a non-Markovian example.

\begin{figure}[t]
\begin{center}
\subfigure[]{\includegraphics[width=3in]{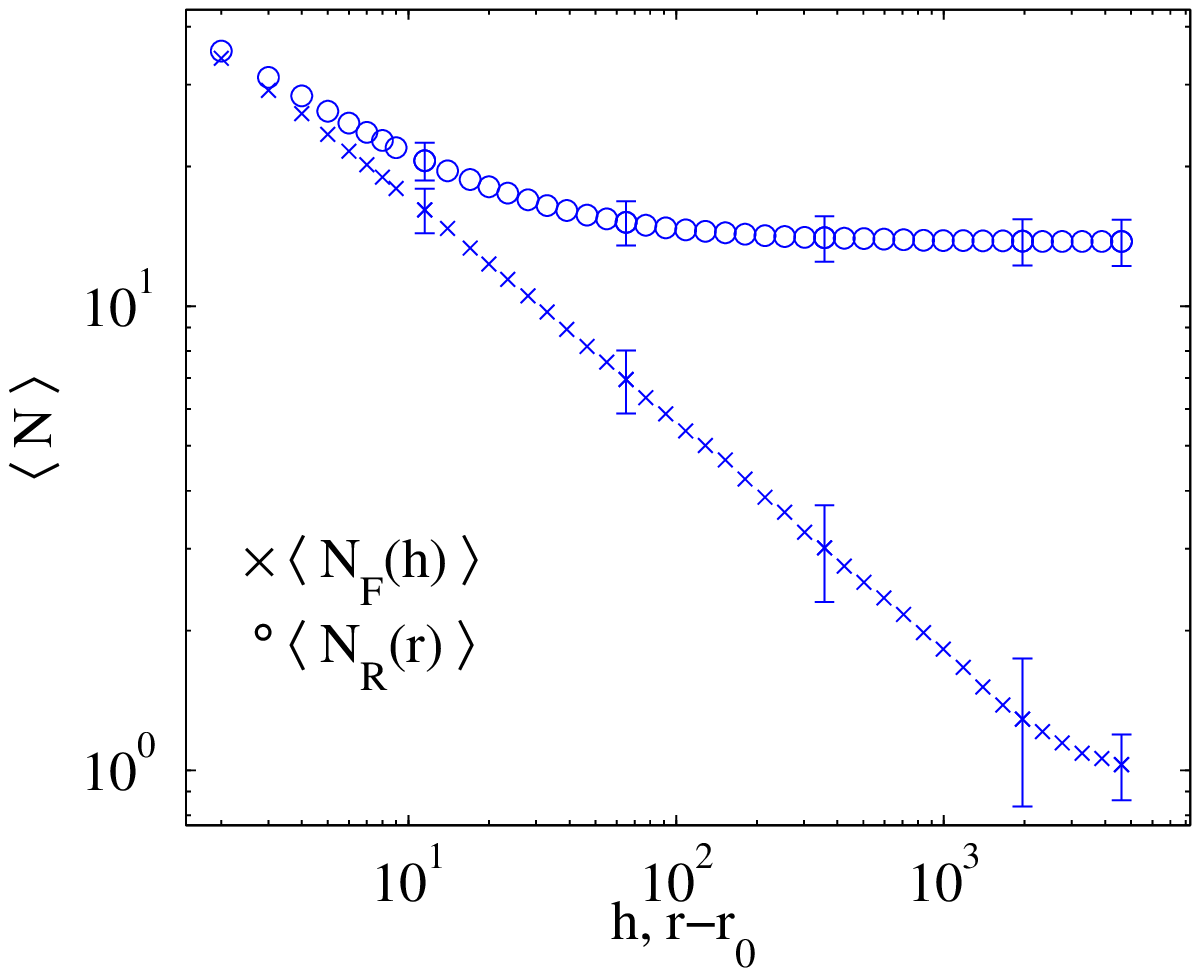}}
\subfigure[]{\includegraphics[width=3in]{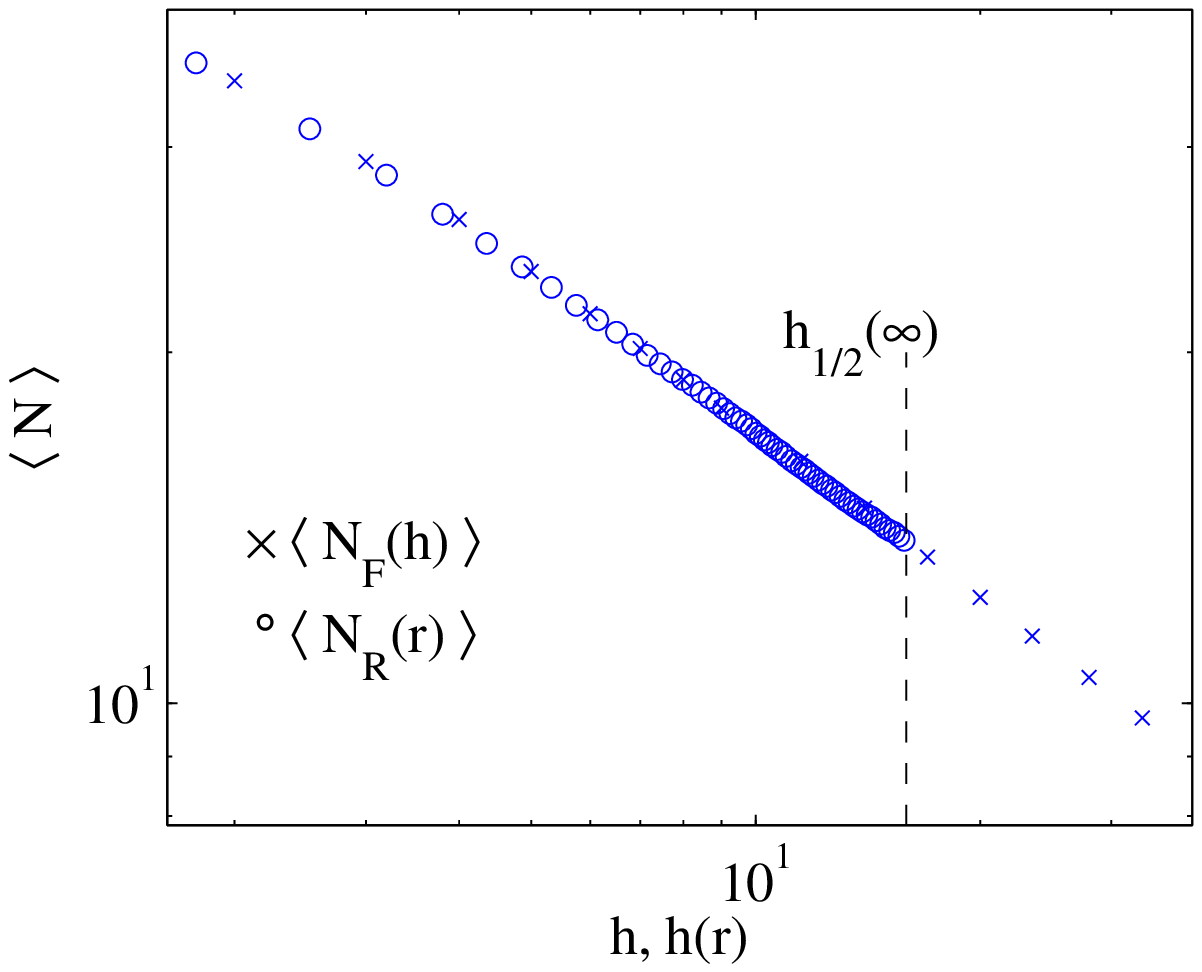}}
\end{center}
\caption{\label{bmref}
Illustration of geometrical effects of expanding domains for coalescing Brownian motions. We compare the average number of particles $\langle N \rangle$ in a fixed domain $[0,L)$ ($\times$) and in a radially expanding domain $[0, 2\pi r)$ ($\circ$) with $r\ge r_{0}$ . For comparison we take $r_{0}=L/2\pi$, where $L=100$ and an initial number of $100$ particles.  
(a) In the fixed domain $\langle N_{F}(h) \rangle$, decreases to the value $1$ corresponding to the absorbing state. However, in the corresponding radially expanding domain $\langle N_{R}(r) \rangle$ decreases to a value greater then $1$. Error bars indicate the standard deviation. (b) Using the mapping (\ref{mapping}) with $\gamma=1/2$, we plot $\langle N_{R}(r) \rangle$ against $h(r)$ and obtain a data collapse. The value $h_{1/2}(\infty)=r_{0}$ is given in (\ref{limit}). 
}
\end{figure}

\subsection{Basic properties of the mapping \label{propmapping}}
Figure~\ref{figmapping} shows the mapping function (\ref{mapping}) for several values of $\gamma$. For $r \rightarrow r_{0}$ we have $h(r) \simeq r-r_{0}$ for all $\gamma>0$, so that initially there is no effect on the particles from the expanding domain, since for $r$ close to $r_{0}$, the fixed and the radial domain are locally equivalent. The non-linear behavior of $h(r)$ encodes the effect of the expanding geometry the large $r$ behaviour is
\begin{equation}
h_{\gamma}(\infty)=\lim_{r\rightarrow \infty} h(r)=\left\{\begin{array}{cl} \frac{\gamma}{1-\gamma}r_{0} &,\ \gamma < 1 \\ \infty &,\ \gamma\ge1 \end{array}\right. \label{limit}.
\end{equation}
The value $h_{\gamma}(\infty)$ corresponds to the height at which the fixed width structure is equivalent to the infinite radius behaviour of the rescaled radially growing structure (see Figure~\ref{bmmap}). This asymptotic behaviour depends on the value of $\gamma$, and for $\gamma<1$ will differ from the analogous asymptotic behaviour in the fixed domain. For fixed domain structures fixation always occurs for interactions such as coagulation or annihilation, i.e. the system will eventually reach an absorbing state, 
as shown in Figure~\ref{bmref}(a). 

Coalescing or annihilating structures in the fixed domain $[0,L)$ are observed e.g. in neutral models for competition in spatial populations (see \cite{Saito1995,Hallatschek2007,Grosskinsky2010,lavrentovich2013} for more details). The fixation time $\tau$ to reach the absorbing state scales with the size $L$ of the system, where by standard arguments
$$\tau \sim L^{1/\gamma} \sim r_{0}^{1/\gamma}.$$
For large systems ($L\rightarrow \infty$), typically $\tau$ is much larger then $h_{\gamma}(\infty)\sim r_{0}$, leading to a non-trivial limit for the statistics of the radial process.
This is because for structures with $\gamma<1$, the spatial expansion rate which is linear in $r$ dominates the lateral spread of random wandering of the particles, where due to the increasing distance, eventually the particles no longer interact. So the statistics for these sub-ballistic structures no longer change and reach an asymptotic value which is random, as indicated by the non-zero and $r$-independent standard deviation in Figure~\ref{bmref}(a). In fact the whole rescaled structure converges to a non-trivial limit where
$$
\Big\{\frac{r_0}{r} Y_r \Big\}
\stackrel{\mathrm{dist.}}{\rightarrow }\big\{ X_{h_{\gamma}(\infty)} \big\} \quad \mbox{as} \quad r\rightarrow \infty.
$$
For structures with $\gamma\ge1$, the particle motion is equivalent to a (super-)ballistic trajectory exceeding the spatial expansion, where from (\ref{limit}) we have $h_{\gamma}(\infty)= \infty $. Here, despite the continuous expansion in space, the asymptotic behaviour for the rescaled radial process will be the same as the analogous behaviour in the fixed domain and we have
$$
\Big\{\frac{r_0}{r} Y_r \Big\}
\stackrel{\mathrm{dist.}}{\rightarrow }\big\{ X_{\infty } \big\} \quad \mbox{as} \quad r\rightarrow \infty.
$$
We can also express the mapping (\ref{mapping}) independently of the system size. Introducing dimensionless variables $r'=r/r_0$ and $h'=h/r_0$ leads to
\begin{equation}\label{smap}
  h'(r')=
  \left\{\begin{array}{cl}
    \frac{\gamma}{1-\gamma}\left( 1-\left( 1/r' \right)^{\frac{1-\gamma}{\gamma}
  }\right)\ &,\ \gamma\neq 1\\
  \log r'\ &,\ \gamma =1\end{array} \right.\ ,
\end{equation}
for all $r'\geq 1$, where for $\gamma =1$ we recover the generic conformal map
from the exterior of the unit circle to a strip. This notation shows that $r_{0}$ plays merely the role of a length scale and $\gamma$ is the only important parameter of the mapping. Moreover, (\ref{smap}) takes the form of a generalized q-logarithm \cite{tsallis2001}, which can therefore be seen as a generic generalization of the standard conformal map in this context.

\section{Applications to self-similar models \label{models}}
In this section, we use the mapping (\ref{mapping}) to characterize radially growing structures as time-rescaled structures in the fixed domain, focusing on coalescence as an example of local interaction. 
We will study the validity of (\ref{mapping}) for self-similar models, for which exact computations are possible. For illustration we show data such as the average number of particles, denoted as $\langle N \rangle $ and the average inter-particle distance squared, denoted as $\langle D^{2} \rangle$ and defined as
\be D^{2}=\sum_{i=1}^{N} (x_{i+1}-x_{i})^{2}. \label{distsq}\ee
Here the particles are ordered such that $x_{i}$ and $x_{i+1}$ are nearest neighbour particle position pairs and the distance is measured modulo periodic boundary conditions. We proceed by considering structures where the particle trajectories are given by three well known self-similar processes, Brownian motion, L\'evy flights and fractional Brownian motion.
We use the representation based on It\^o calculus, since this is most convenient to describe the effects of time changes which are vital for our analysis. There are also well developed extensions of It\^o calculus to fractional Brownian motion and L\'evy processes which we will make use of.

\subsection{Brownian motion \label{Bm}}
Consider a standard Brownian motion $B:= (B_{t},t\ge 0)$, which is a centered Gaussian process with mean $\langle B_{t} \rangle =0$ and covariance
$$\langle B_{t}B_{s} \rangle =\mathrm{min}(t,s)\ ,$$
and is self similar and locally scale invariant with $\gamma =1/2$. 
In this case we can give a rigorous derivation of the mapping $h(r)$ using It{\^o}-calculus. Consider the arms $X:= (X_{h},h\ge 0)$ and $Y:= (Y_{r},r\ge r_{0})$ as defined in Section~\ref{secresults}. The corresponding Langevin equations in standard notation of stochastic differential equations are \cite{Oksenda2003}, \be\label{dX} dX_{h}=dB_{h}\ee and \be\label{dY}
  dY_r = Y_r\,dr/r + dB_r.
\ee
The first term in (\ref{dY}) corresponds to the stretching of space which is proportional to the angular displacement $Y_{r}/r$, the second term in (\ref{dY}) describes the
inherent fluctuations of the process, 
where $dB$ denotes infinitesimal increments of a standard Brownian motion. For the rescaled process
\be \label{rescaledp}
 Z:= (Z_{r},r\ge 0) \quad \mbox{with}\quad Z_r =\frac{r_0}{r}\, Y_r \,,
\ee
using It{\^o}'s formula we have
\be
\label{dZ}
  dZ_r =\frac{ r_{0}}{r}\,dB_{r}.
\ee
In integral form this implies  $Z_r = \int_{r_{0}}^{r}r_{0}/s \,dB_{s}.$
It is well known that this stochastic It\^o integral can be written as a time-changed Brownian motion so that the process $Z$ is a continuous time martingale, where $Z_r = B_{h(r)}$. The time change is given by the quadratic variation of $Z$
\be
h(r)=\int_{r_{0}}^{r}\Big(\frac{r_{0}}{s}\Big)^{2}\,ds =r_0 \Big( 1-\frac{r_0}{r}\Big)\ .\label{hr2}
\ee
The quadratic variation basically describes the accumulated mean squared displacement up to time $t$, and is simply equal to $t$ for standard Brownian motion. In general, every It\^o integral with respect to Brownian motion such as $Z_r$ is a martingale, and can be written as a time-changed Brownian motion using the above formula (see e.g. \cite{Oksenda2003} page 56 for details).
This gives a rigorous justification of the mapping (\ref{mapping}) for Brownian motion with $\gamma=1/2$ and equivalence in the strong sense (\ref{strongres}), 
i.e. the full process including time correlations is mapped correctly.

\subsection{L\'evy flights \label{levymodels}}
Structures where the position of the arms exhibit super-diffusive behaviour due to jumps in their trajectories can be modelled by $\alpha$-stable L\'evy processes \cite{Manuel1999,Fogedby1994}. A stochastic process $L=(L^{\alpha}_{t}, t\ge 0)$, with $\alpha \in (0,2)$ is an $\alpha$-stable L\'evy process if it has stationary independent increments and a pdf $p_{\alpha}(x,t)$ whose Fourier transform takes the form 
\begin{equation} \hat{p}_{\alpha}(k,t)=\exp(-\sigma_{\alpha}t|k|^{\alpha}/2). \label{levypdf} \end{equation} 
L\'evy Processes are Markovian with discontinuous paths, they have increments with infinite variance, and for $\alpha \in (0,1)$ their absolute first moment is also infinite.
Another key property is that the fractional moments of $L^{\alpha}_{t}$ scale as \be \label{qproperty} \langle |L^{\alpha}_{t}|^{q}\rangle = (\sigma_{\alpha} t^{1/\alpha})^{q}, \ee where $0<q<\alpha$ is non-integer \cite{Chechkin2008,Manuel1999}.
In practice, the parameter $\alpha$ can be greater then $2$, but in this case the increments have a finite mean and variance, and such a process scales diffusively.

When the arms $X$ and $Y$ are $\alpha$-stable L\'evy processes, through a generalization of It\^o calculus, \cite{Chechkin2008,Applebaum2009}, the rescaled process $Z$ can be expressed in Langevin form as
\be
\label{dZa}
  dZ_r =\frac{ r_{0}}{r}\,dL^{\alpha}_{r}.
\ee 
The process $Z$ is also a time-changed $\alpha$-stable L\'evy process (see \cite{Applebaum2009}, page 237). By using the property (\ref{qproperty}) on (\ref{dZa}) we can match all $q$-moments of $dX$ and $dZ$, 
$$ (c_{\alpha} dh)^{q/\alpha}= \Big(\frac{r_{0}}{r}\Big)^{q} (c_{\alpha} dr)^{q/\alpha}, $$
provided that
$$
  h(r) = \int_{r_{0}}^{r}\Big(\frac{r_{0}}{s}\Big)^{\alpha}\,ds\ ,
$$
which is the mapping (\ref{mapping}) with $\gamma =1/\alpha$. Therefore the process $Z_{r}$ is an $\alpha$-stable L\'evy process $L^{\alpha}_{h(r)}$ and the result holds in its strong form (\ref{strongres}). Note that for $\alpha\geq 2$ L\'evy processes scale like Brownian motion and do not become subdiffusive, so that for general $\alpha >0$ we have $\gamma=\mathrm{max}\{1/\alpha,1/2\}$.

\begin{figure}[t]
\begin{center}
\subfigure[]{\includegraphics[width=3in]{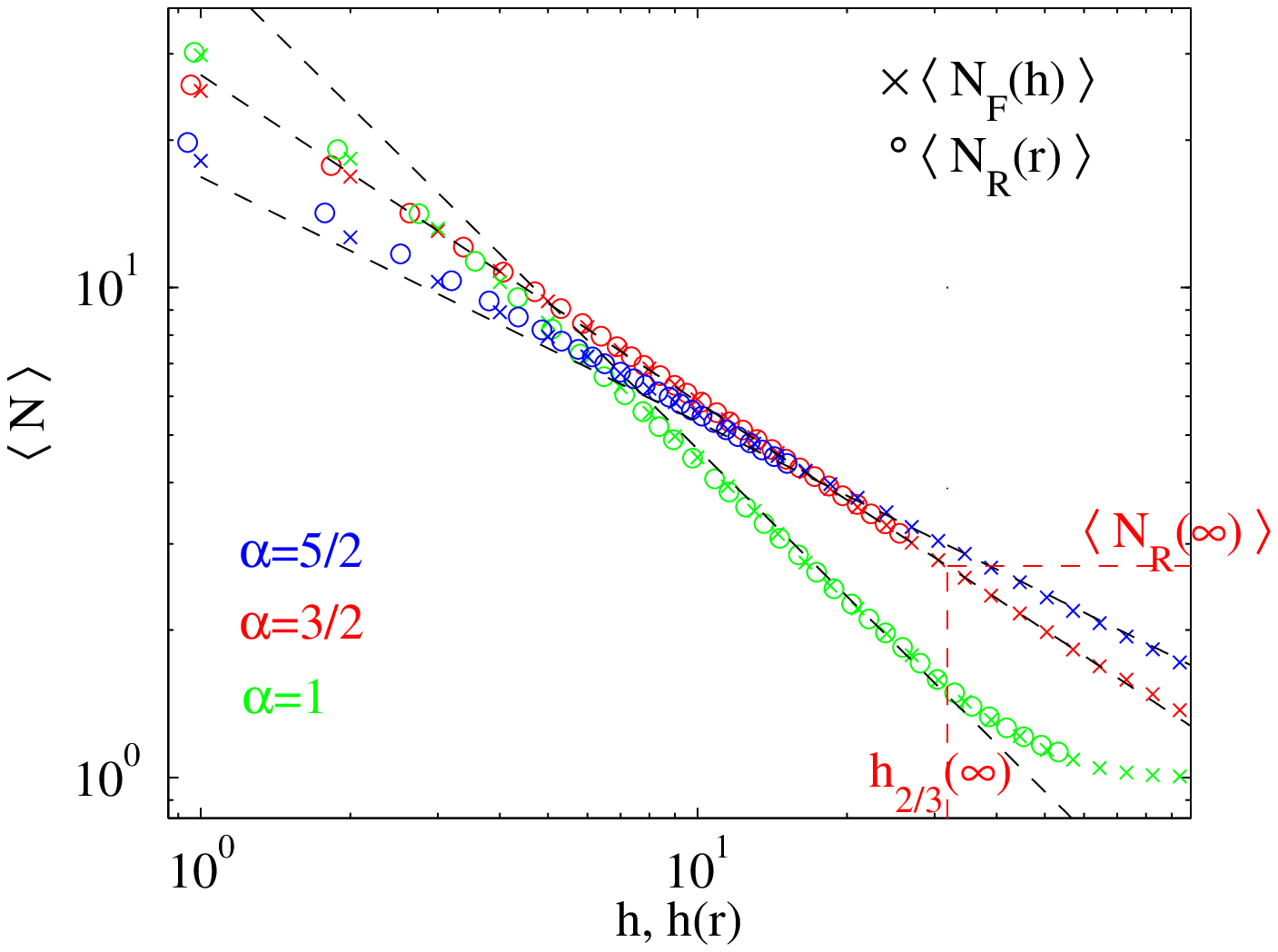}}
\subfigure[]{\includegraphics[width=3in]{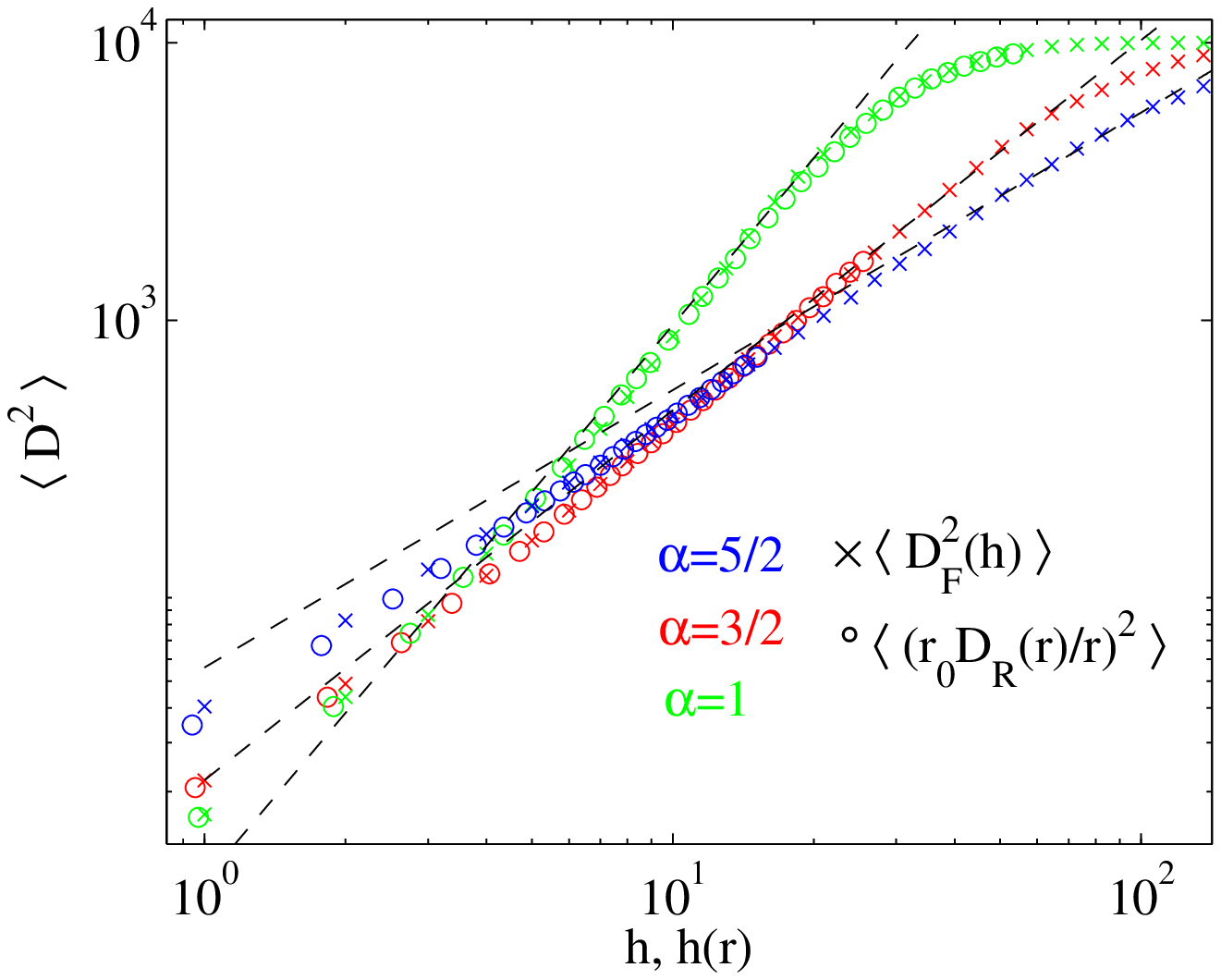}}
\end{center}
\caption{\label{levystats}
Numerical confirmation of the mapping (\ref{mapping}) between radial
geometry ($\circ$) and fixed domain ($\times$) for $\alpha$-stable L\'evy flights (\ref{levypdf}) with $\gamma=\mathrm{max}\{1/\alpha,1/2\}$. Data is gathered for $L=100$ with an initial number of $100$ particles and $r_{0}=L/2\pi$. By plotting the radial data against $h(r)$ we map (a) $\langle N_{R} \rangle $ and (b)  $\langle (r_{0}D_{R}(r)/r)^{2} \rangle$  to the fixed data. Dashed lines indicate mean field results (see \ref{ss2A1}) and the asymptotic value for the radial domain (\ref{Nrlimit}) is illustrated for $\alpha=3/2$ in (a).} 
\end{figure}

Figure~\ref{levystats}(a) shows the expected number of particles $\langle N_{F}(h) \rangle$ and $\langle N_{R}(r) \rangle$, plotted against $h$ and $h(r)$, respectively, for several values of $\alpha$ under coalescing dynamics. The figure shows that we have a very good data collapse, where the radial data converges to $\langle N_{F}(h_{\gamma}(\infty)) \rangle$ as explained in Section~\ref{propmapping}. Being Markov processes with independent increments, we can simulate L\'evy processes simply by adding independent increments. We take the absolute value of each increment $dL^{\alpha}$ to be distributed by the Pareto distribution with pdf \be p_{\alpha}(x)=\alpha b^{\alpha}/x^{\alpha+1} \quad \mbox{for} \quad x \ge b, \label{pareto}\ee where $b=dt^{1/\alpha}$ with $dt$ being the simulation time-increment. Taking $dt\ll 1$ small we reproduce data which is consistent with the infinitesimal limit $dt\rightarrow 0$, and therefore we simulate good approximations of sample paths of $\alpha$-stable L\'evy processes \cite{Chechkin2008}. Coalescence or annihilation interaction can easily be implemented even for discontinuous paths with discrete time sampling, since in one space dimension every event can be detected by a change of the particle order. We do not include multiple coalescence events, i.e. if a particle jumps across several others, we just coalesce it with the nearest one at the position of the latter.

The long time behaviour in the fixed domain can be computed analytically (see \ref{ss2A1}), where in a scaling window of intermediate $h$
\be \langle N_{F}(h)\rangle= L/(\pi \sigma_{\alpha} h^{2/\alpha})^{1/2} \label{Nlevy},\ee
and this is shown by the black dashed line in Fig.~\ref{levystats}a.
Using (\ref{Nlevy}) and (\ref{limit}) the corresponding asymptotic behaviour for the radial structure is
\be
\langle N_{R}(r)\rangle \rightarrow \left\{\begin{array}{cl} 2\sigma_{\alpha}^{-1} \pi^{1/2} \Big( \frac{r_{0}^{1-\alpha}}{\alpha-1}\Big)^{-1/\alpha} &,\ \alpha > 1  \\ 1 &,\ \alpha \le 1 \end{array}\right. \quad \mbox{as} \quad r\rightarrow \infty \label{Nrlimit}
\ee
and this limit is indicated for the $\alpha=3/2$ data by the red dashed lines, where the values for $\sigma_{\alpha}\approx 1.4$, $4$ and $10.8$ for $\alpha=1,$ $3/2,$ and $5/2$ respectively, are fitted to the data.

Figure~\ref{levystats}(b) shows the corresponding behaviour of $\langle D^{2} \rangle$ as defined in (\ref{distsq}). Both $\langle D_{F}^{2} \rangle$ and the $\langle D_{R}^{2} \rangle$ are increasing functions, and in the fixed domain $\langle D_{F}^{2}(h) \rangle$ will converge to $L^{2}$. For the radial domain the rescaled behaviour $\langle(r_{0} D_{R}(r)/r)^{2}\rangle$ is shown, where by plotting against $h(r)$ we attain a data collapse.

\subsection{Fractional Brownian motion}
We also consider structures where the displacement of the arms perform fractional Brownian motion (fBm) \cite{Biagini2008,Hahn2011}. The fBm $B^{H}=(B^{H}_{t}, t \ge 0)$ with Hurst exponent $H\in (0,1)$ is a centered Gaussian process with continuous paths and covariance
\be\label{covma}
\langle B^{H}_{t}B^{H}_{s} \rangle =\frac{1}{2} (t^{2H} +s^{2H} -|t-s|^{2H})\ .
\ee
In particular, $\langle (B^{H}_{t})^2 \rangle =t^{2H}$, the pdf of $B_t^H$ takes the general form
\begin{equation}p(x,t)=\frac{1}{\sqrt{2\pi t^{2H}}}\exp\Big(-\frac{x^{2}}{2t^{2H}}\Big)\label{fbmpdf} \end{equation}
and the process exhibits local scale invariance (\ref{locsca}) with $\gamma=H$. When $H=1/2$, the process is a Brownian motion as in Section~\ref{Bm}, and  for $H\ne1/2$ the process is not Markovian since it has long range temporal correlations \cite{Biagini2008}. 
When the arms are fBm as before the rescaled process (\ref{dZ}) can be written in integral form as

\be \label{Zfbm}
  Z_r = \int_{r_{0}}^{r}\frac{r_{0}}{s}\,dB^{H}_{s}.
\ee
For $H \ne 1/2$ this integral w.r.t fBm cannot be written as a time-changed fBm \cite{Hu2007}, so the mapping does not hold in its strong form (\ref{strongres}). This is due to memory effects coming from the non-Markovian correlated noise $dB_{t}^{H}$ leading to non-independent increments. Nevertheless, using fractional calculus we can compute and match the second moment of the rescaled radial process and the fixed process $B^{H}$. We represent $Z_{r}$ in (\ref{Zfbm}) as a memory kernel integral with respect to a standard Brownian motion (see \cite{Biagini2008} page 48), leading to the following representation:
\be Z_{r} = \int_{r_{0}}^{r} \Big(K^{*}_{H} \frac{r_{0}}{(\cdot)}\Big)(s)dB_{s} \label{fbmsdebm}.\ee
The operator $K_{H}$ and further details are given in \ref{asde}. In the form given in (\ref{fbmsdebm}), we match the quadratic variations of $Z_r$ with $X_{h_{H}(r)}$ to obtain
\be h_{H}(r)= \Big[H(2H-1)\int_{r_{0}}^{r} \int_{r_{0}}^{r} \frac{r_{0}^2}{xy} |x-y|^{2H-2} dxdy\Big]^{1/2H} \label{mappingfbm1}. \ee
This can also be written using hypergeometric functions 
\begin{eqnarray}
h_{H}(r) = \Bigg[Hr_{0}^{2}\int_{0}^{r-r_{0}}\Big[ &\frac{(r-r_{0}-y)^{-1+2 H} {}_{2}F_{1}\left[1,-1+2 H,2 H,\frac{r_{0}-r+y}{r_{0}+y}\right]}{(r_{0}+y)^{2}} + \nonumber \\
 &\frac{y^{-1+2 H} {}_{2}F_{1}\left[1,1,2 H,-\frac{y}{r_{0}}\right]}{r_{0}(y+r_{0})} \Big] dy\Bigg]^{1/2H} \label{mappingfbm2} \,.
\end{eqnarray}

\begin{figure}[t]
\begin{center}
\subfigure[]{\includegraphics[width=3in]{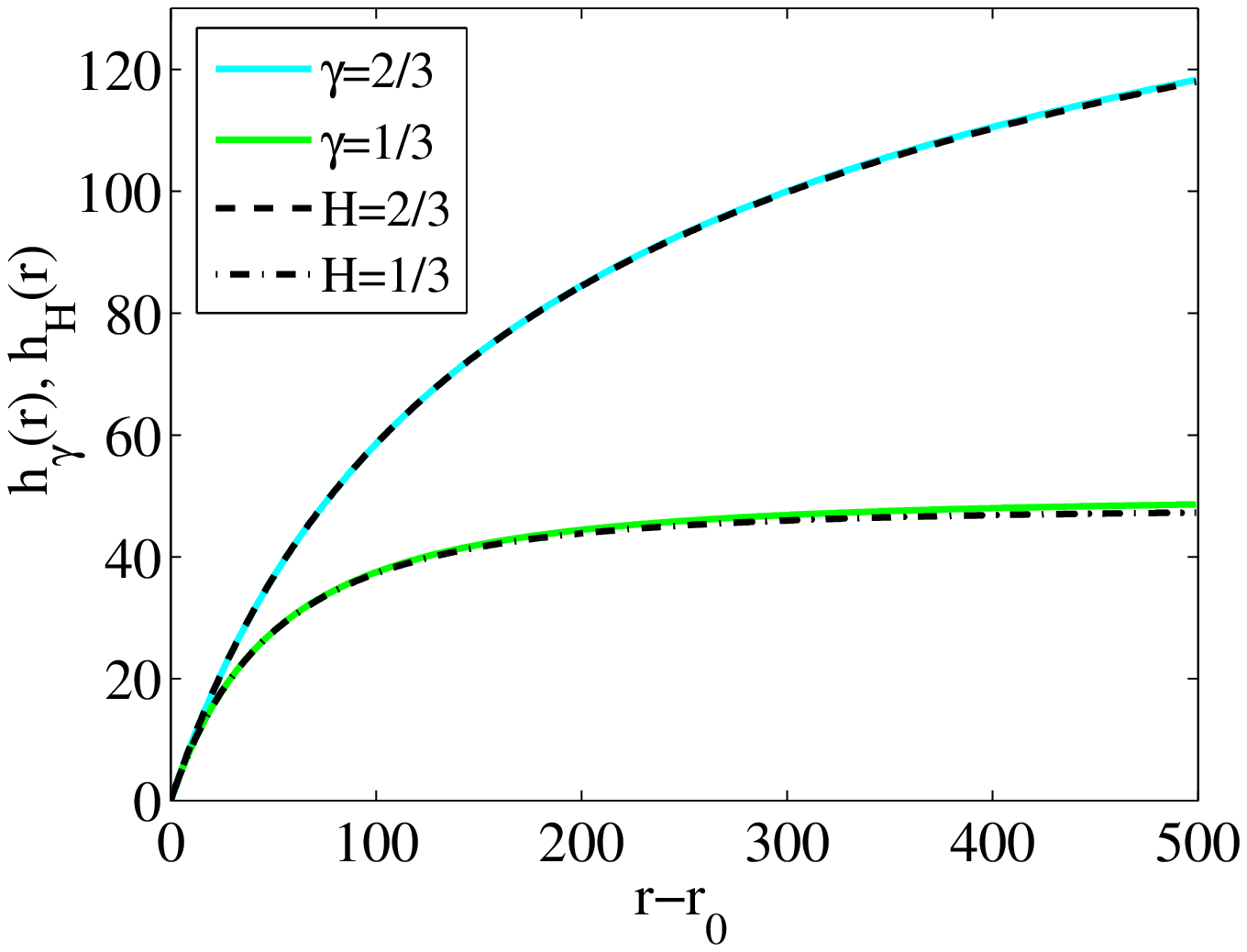}}
\subfigure[]{\includegraphics[width=3in]{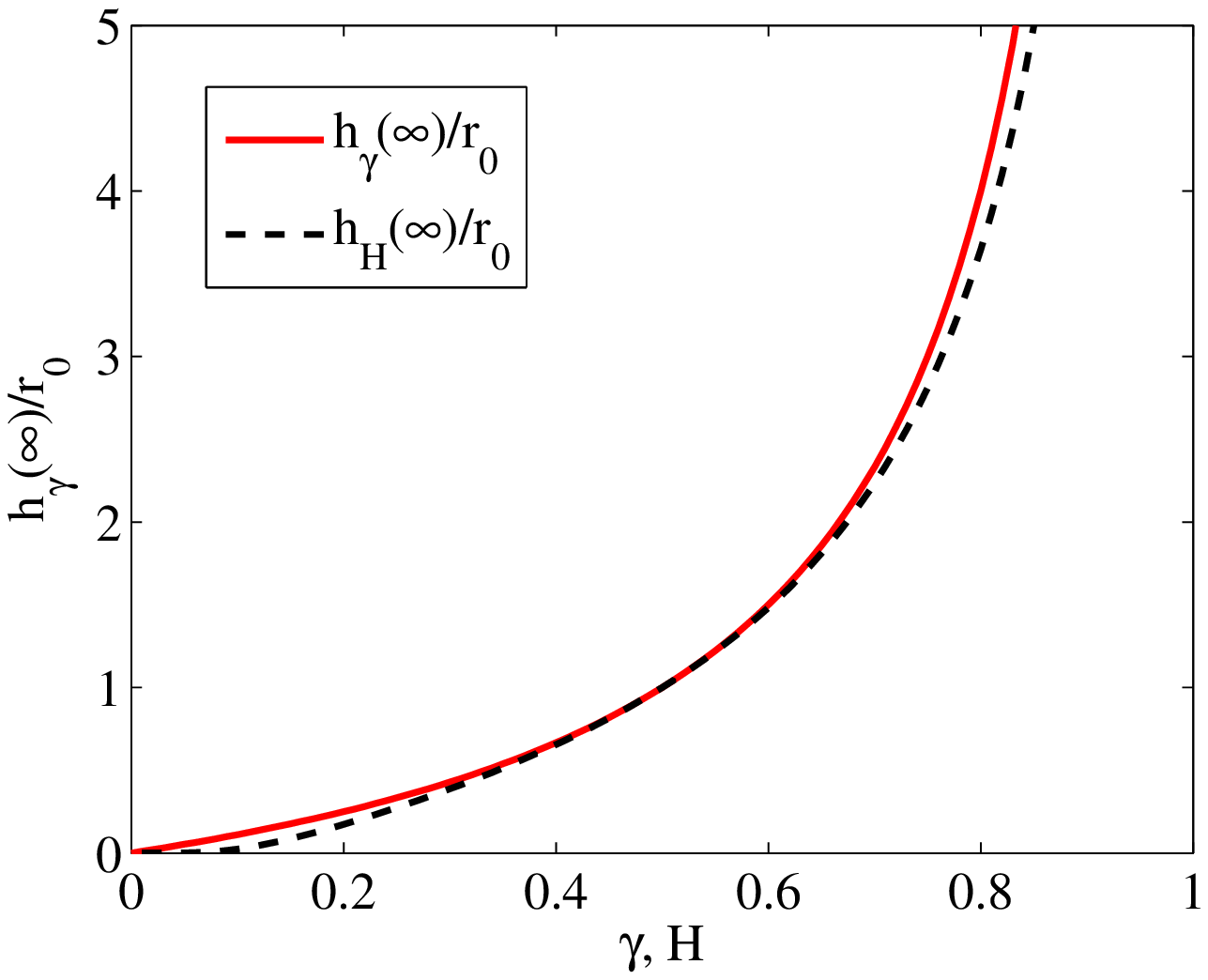}}
\end{center}
\caption{\label{figfbmmapping1}
Comparison of the mapping (\ref{mapping}) (full color) and (\ref{mappingfbm1}) (dashed, black) for $H=\gamma$ and $r_{0}=100$. (a) Both curves are very similar and behave as $r-r_{0}$ for $r$ close to $r_{0}$. For large $r$ the functions differ, as shown in (b) the limit $h_{\gamma}(\infty)$ (\ref{limit}) and $h_{H}(\infty)$ (\ref{fbmlimit2}) do not match, but are very close.
}
\end{figure}

\begin{figure}[t]
\begin{center}
\subfigure[]{\includegraphics[width=3in]{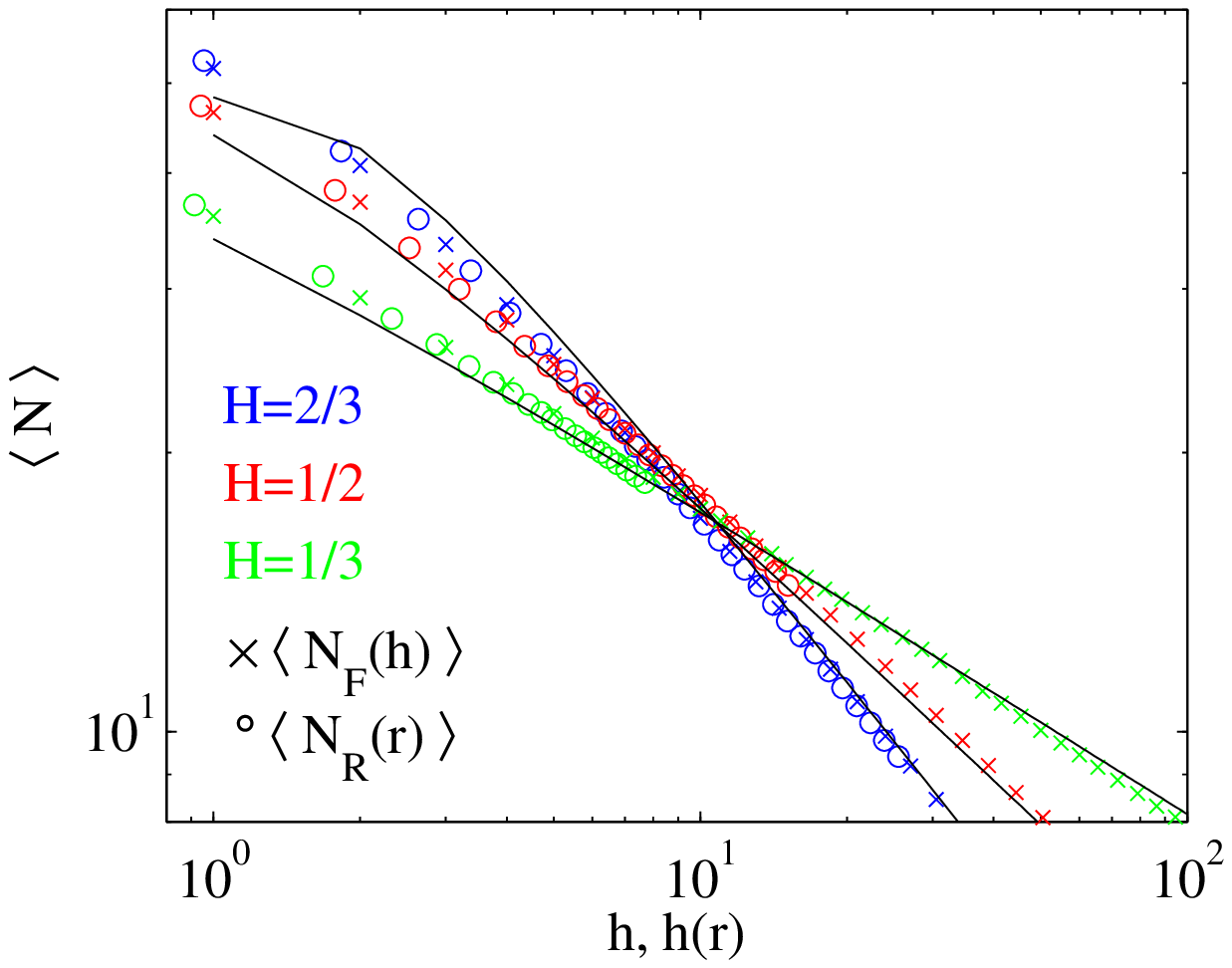}}
\subfigure[]{\includegraphics[width=3in]{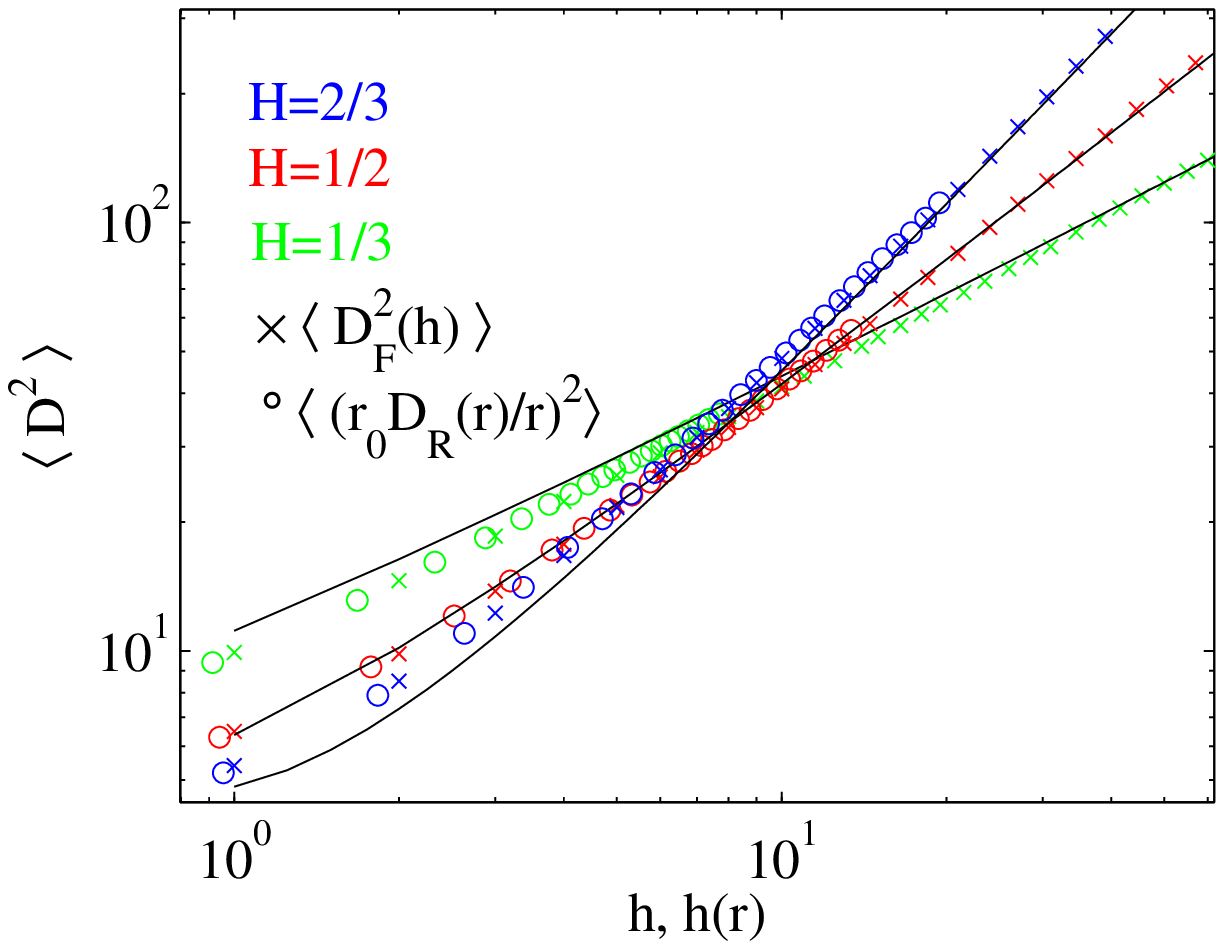}}

\end{center}
\caption{\label{fbmstats}
Numerical confirmation of the mapping (\ref{mapping}) between radial
geometry ($\circ$) and fixed domain ($\times$) for fBm structures with $H=\gamma$. Data is gathered for $L=100$ and $r_{0}=L/2\pi$ with an initial number of $100$ particles. Plotting (a) $\langle N_{R} \rangle $ vs $h(r)$, and (b) $\langle (r_{0}D_{R}(r)/r)^{2} \rangle$ vs $h(r)$ gives a good data collapse. 
Solid lines indicate exact predictions (see \ref{ssA1}). }

\end{figure}

This second representation also holds for $H=1/2$, where it simplifies to (\ref{hr2}). Note that although for $H\ne1/2$ the expression of (\ref{mappingfbm1}) differs from (\ref{mapping}), visual comparisons as plotted in Figure~\ref{figfbmmapping1}(a) for $H=1/3$ and $H=2/3$ shows that they are very close.
In fact, for the rest of this paper (especially Figs.~\ref{fbmstats} and
\ref{cdwalkers}) we do not distinguish between the two mappings.
However, for fBm we are only able to match first and second moment of the processes $Z_r$ and $X_{h(r)}$, and the equivalence in the form (\ref{strongres}) does not hold. Since Gaussian random variables are uniquely determined by their first and second moment, we still have
$$
\Big\{\frac{r_0}{r} Y_r \Big\}
\stackrel{\mathrm{dist.}}{=}\big\{ X_{h(r)} \big\} \quad \mbox{for all} \quad r\geq r_0 \ ,
$$
i.e. equivalence for all marginals but not for the time correlations. 
The behaviour for small radius $r \in [r_{0}, r_{0}+\epsilon]$ using (\ref{mappingfbm1}) is
\begin{eqnarray}
\lim_{\epsilon \rightarrow 0} \frac{h_{H}(\epsilon+r_{0})}{\epsilon} &=& \lim_{\epsilon \rightarrow 0} \bigg[ H(2H-1)\int_{0}^{1}\int_{0}^{1} \frac{r_{0}^{2}|x-y|^{2H-2}}{(r_{0}+\epsilon x) (r_{0}+\epsilon y) } dxdy \bigg]^{1/2H} \nonumber \\
&=&\Big[ H(2H-1)\int_{0}^{1}\int_{0}^{1} |x-y|^{2H-2}dxdy \Big]^{1/2H}=1.\nonumber
\end{eqnarray}
So $h(r) \simeq r-r_{0}$ for $r$ close to $r_{0}$.
Note that (\ref{mappingfbm1}) and (\ref{mapping}) will mostly differ when $r$ is large (see Figure~\ref{figfbmmapping1}(b)), where in the limit as $r\rightarrow \infty$ we have
\begin{eqnarray}
\lim_{r \rightarrow \infty} h_{H}(r)&=& r_{0}\Big[ H(2H-1)\int_{0}^{\infty}\int_{0}^{\infty} \frac{|x-y|^{2H-2}}{(x+1) (y+1) } dxdy \Big]^{1/2H} \nonumber \\
 &=& r_{0}\pi ^{\frac{1}{2 H}} \left(\frac{H (2 H-1) }{(H-1)\sin(2\pi H)}\right)^{\frac{1}{2 H}} . \label{fbmlimit}
\end{eqnarray}
Comparing this value to (\ref{limit}) we have
\be h_{H}(\infty) =\lim_{r \rightarrow \infty} h_{H}(r)\quad \left\{\begin{array}{cl} =h_{\gamma}(\infty)=\frac{\gamma}{1-\gamma}\, r_{0} &,\ H =1/2   \\ < h_{\gamma}(\infty)=\frac{\gamma}{1-\gamma}\, r_{0} &,\ H\ne 1/2 \end{array}\right. . \label{fbmlimit2}
\ee

Figure~\ref{fbmstats} shows the statistics $\langle N \rangle$ and $\langle D^{2} \rangle$ for radially expanding and fixed fBm structures. Since fBm is non-Markovian, sample paths cannot be generated by adding increments that depend only on the current state. The easiest way to generate a discretized sample of an fBm path at times $t_1 ,\ldots ,t_n$ which we implemented, is to invert the covariance matrix $\langle B^{H}_{t_i}B^{H}_{t_j} \rangle$ given by (\ref{covma}) and multiply with a vector of iid Gaussian increments (cf. \cite{Biagini2008} Section 10 for details). For the radially expanding structure we use the mapping (\ref{mapping}) with $\gamma=H$ and plot $\langle N_{R}(r) \rangle$ vs $h(r)$. This gives a good data collapse despite as shown above this is strictly not the correct mapping. 
Figure~\ref{fbmstats}(b) shows the behaviour of $\langle D^{2} \rangle$. This follows similar behaviour to the L\'evy data [as seen in Figure~\ref{levystats}(b)]. Again by plotting the rescaled $\langle (r_{0}D_{R}(r)/r)^{2} \rangle$ against $h(r)$ we have a data collapse. For fBm the full behaviour of $\langle N_{F}(h) \rangle$ and $\langle D_{F}(h)^{2}\rangle$ can be analytically approximated (see \ref{ssA1}) which is shown by the full black lines.

\subsection{More general self-similar processes}

As we have seen before, the mapping (\ref{mapping}) is unique for all locally scale invariant Markovian models. However, there are many other self-similar systems which are not locally scale invariant, for which similar but different mappings can be derived rigorously. Here we focus on a special example of Brownian motion with 
time dependent diffusion coefficient. For such processes the equations governing the fixed and radial process are
\be\label{dXg} dX_{h}=\phi(h) dB_{h}\ee
and
\be
  dY_r = Y_r\,dr/r + \phi(r-r_{0})dB_r.
\ee
Here $B$ is taken to be a standard Brownian motion with exponent $\gamma=1/2$. The function $\phi: [0,\infty) \rightarrow [0,\infty)$ is a continuous positive function, some interesting forms of $\phi$ are power laws such that $\phi(t)\sim t^{\delta}$. Such a function can be seen to occur for biologically motivated models as in \cite{Grosskinsky2010} and are also observed in DLA structures \cite{Ali2013}. In this case we have
$$
\langle X_h^2 \rangle \sim \int_0^h s^{2\delta} ds\sim h^{2\delta +1} \ ,
$$
and we consider $-1/2 <\delta <1/2$ which leads to sub- and superdiffusive processes with Hurst exponent $\frac12 +\delta$.
Treating these processes as before, the rescaled process (\ref{rescaledp}) can be seen to satisfy
\be
\label{dZg}
  dZ_r =(r-r_{0})^{\delta} \frac{r_{0}}{r}\,dB_{r}\ .
\ee
Taking the fluctuations in (\ref{dXg}) and (\ref{dZg}) to satisfy (\ref{locsca}) and using the fact that the rescaled process has the same law as the process in the fixed domain leads to the following relationship between the temporal coordinates $(h,r)$

$$h_{\delta}(r)=\left[(1+2\delta)\int_{r_{0}}^{r} \Big(\frac{r_{0}}{s}\Big)^{2}(s-r_{0})^{2\delta}\, ds\right]^{\frac{1}{2\delta+1}} \ .$$
This can also be written as
\begin{eqnarray}
h_{\delta}(r)&= \Bigg[(1+2\delta)r_{0}^{2}\Gamma\Big(1-2\delta\Big) \Bigg[r_{0}^{-1+2\delta}\, \Gamma\Big(2\delta+1\Big)-\nonumber \\
 &\qquad {} r^{-1+2\delta}\, {}_{2}F_{1}\left[-2\delta,1-2\delta,2-2\delta,\frac{r_{0}}{r}\right]\Bigg]\Bigg]^{1/(2\delta+1)} \label{mappingg},
\end{eqnarray}
 where $\Gamma(\cdot)$ is the gamma function. Thus the mapping (\ref{mappingg}) complements the one for locally scale invariant structures (\ref{mapping}), applying to a different class of self-similar processes and resulting stochastic structures. For $-1/2<\delta<1/2$ (\ref{mappingg}) has the finite limit

\be h_{\delta}(\infty)= \Bigg[(1+2\delta)r_{0}^{2\delta+1} \Gamma\Big(1-2\delta\Big) \Gamma\Big(1+2\delta\Big) \Bigg]^{1/(1+2\delta)}. \label{limitnl}\ee

\begin{figure}[t]
\begin{center}
\subfigure[]{\includegraphics[width=3in]{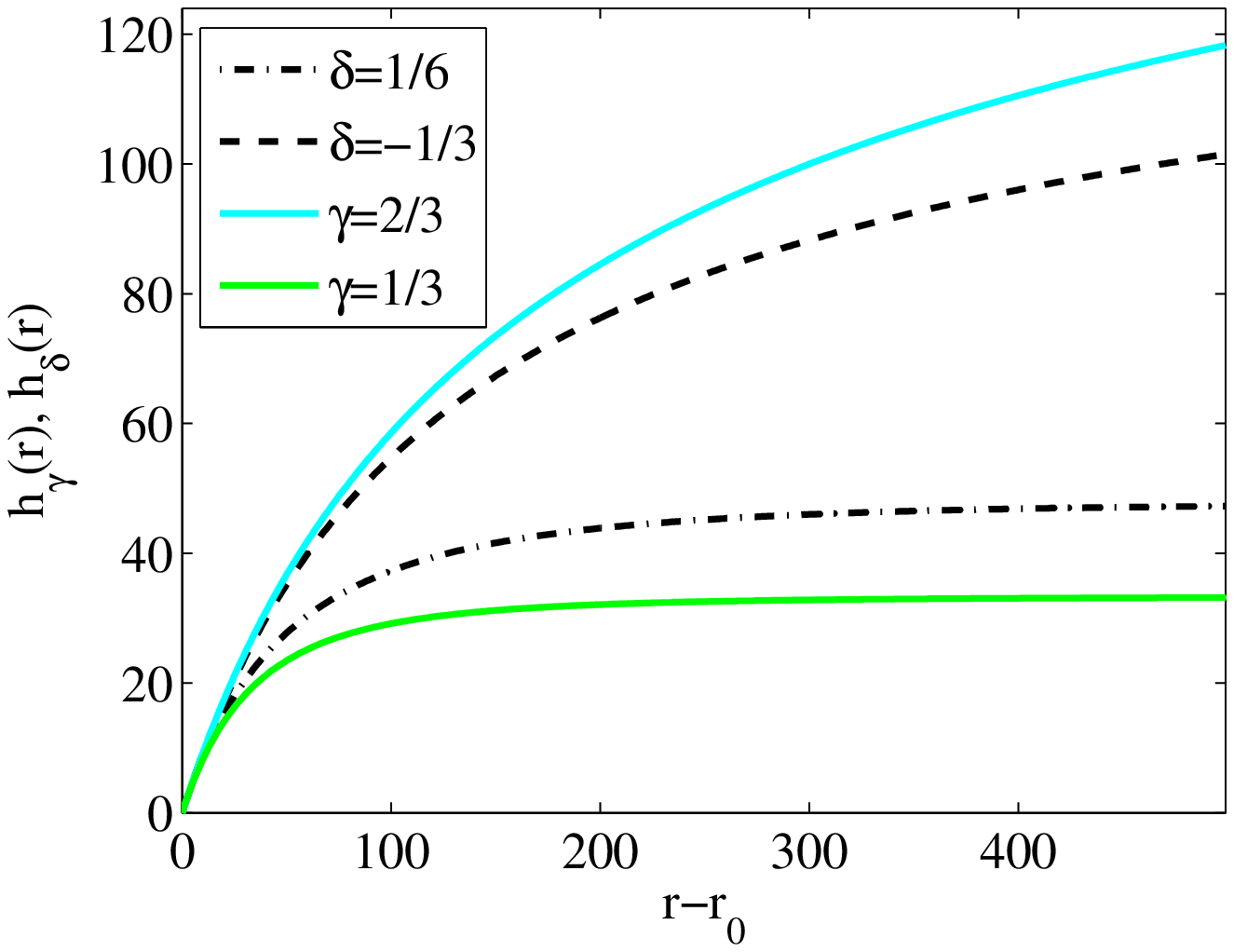}}
\subfigure[]{\includegraphics[width=3in]{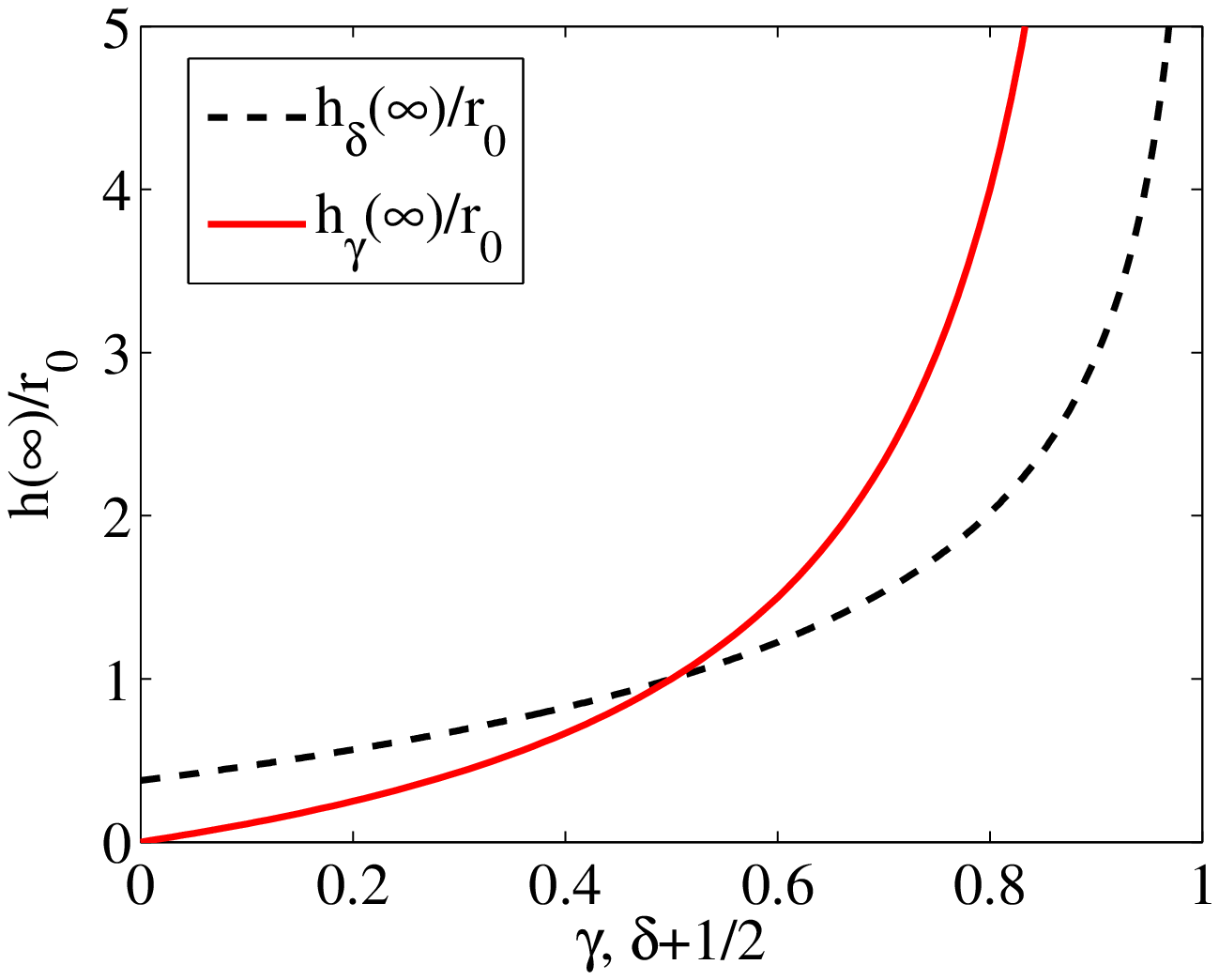}}
\end{center}
\caption{\label{figfbmmapping2}
Comparison of the mapping (\ref{mappingg}) (full color) and (\ref{mappingfbm1}) (dashed, black) for $\gamma=1/2+\delta$ and $r_{0}=100$. (a) Both curves have the same monotonic feature, the mapping $h_{\delta}(r)$ behaves as $r-r_{0}$ for $r$ close to $r_{0}$, and deviates from $h_{\gamma}(r)$ for $r$ large. (b) The limit $h_{\gamma}(\infty)$ (\ref{fbmlimit2}) and $h_{\delta}(\infty)$ (\ref{limitnl}) do not match, only for $\delta _0$ and $\gamma =1/2$.
}
\end{figure}

In Figure~\ref{figfbmmapping2} we compare both mappings (\ref{mapping}) and (\ref{mappingg}).
We take the parameter $\delta$ such that $\gamma =1/2 +\delta$, this equates the exponent of the mean squared displacement for each arm in their respective system. In Figure~\ref{figfbmmapping2}(a) we see that despite initial similarity, for larger values of $r$ both mappings converge to clearly different limits, as is illustrated in Figure~\ref{figfbmmapping2}(b). This difference is far clearer than differences between (\ref{mapping}) and the mapping for fBm processes (\ref{mappingfbm1}), and shows that processes with the same mean squared displacements cannot necessarily be mapped by similar functions. Note also that by construction $X_h$ (\ref{dXg}) and the rescaled process $Z_r$ (\ref{dZg}) are time-changed Brownian motions in this example, and therefore the mapping holds in its strong form (\ref{strongres}).

\section{Generalized geometries \label{generaldomain}}
In Section~\ref{secmapping} we have shown how radially growing structures can be mapped to structures growing on the fixed domain. In this section we generalize our theory by considering evolution on a general time dependent, isotropic domain.

\subsection{Decreasing radial domain}
\begin{figure}[t]
\begin{center}
\subfigure[]{\includegraphics[bb=122 287 490 577,clip,width=2in,height=2in]{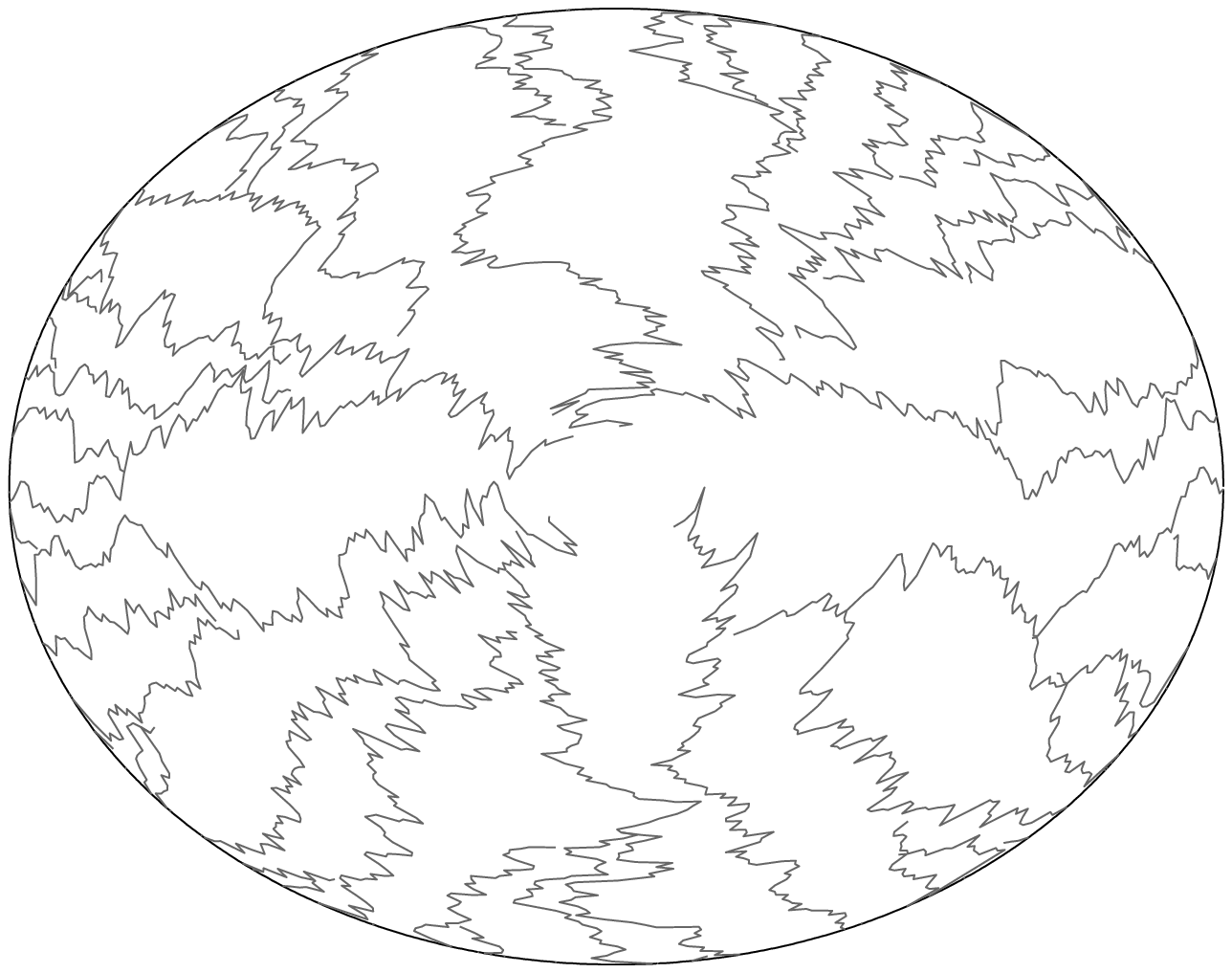}}
\subfigure[]{\includegraphics[width=3in]{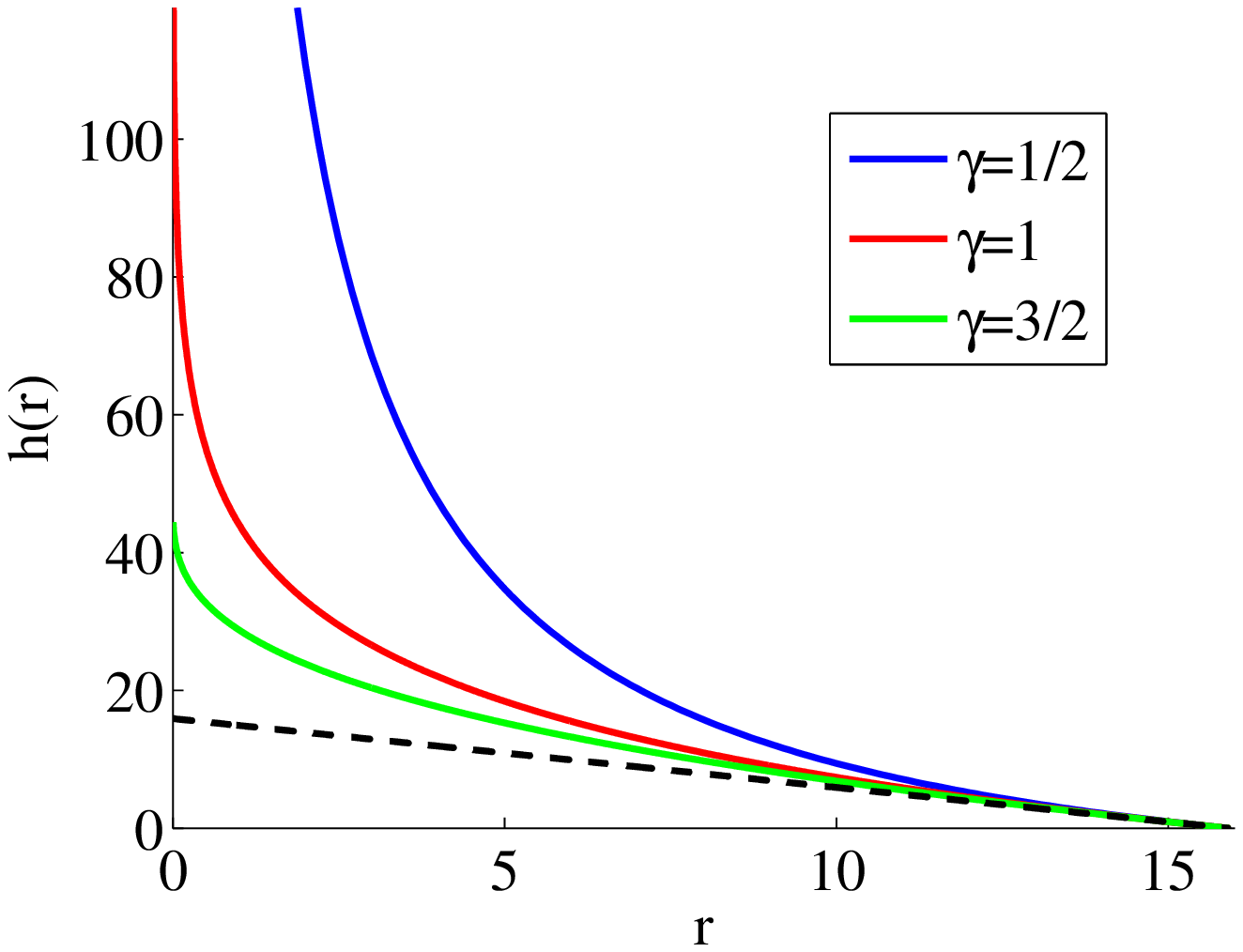}}
\end{center}
\caption{\label{bminside}
(a) Radially decreasing coalescing Brownian structure with $r_{0}=100/(2\pi)\approx 15.9$ and $100$ initial particles. Here each particle performs a directed path inwards.
(b) The inward mapping (\ref{mappingcd}) for $r_{0}\approx 15.9$ and several values of $\gamma$. Analogous to the outward mapping (\ref{mapping}) the initial behaviour is linear where $h(r) \simeq r_{0}-r$ (black dashed line). The asymptotic behaviour is dependent on $\gamma$, where for $\gamma>1$ the mapping has a finite limit. Despite this finite value all inward growing structures will fixate, as is explained in the text.
}
\end{figure}

We start by considering decreasing radial structures where the length of the domain decreases uniformly as a function of the radius. This particular geometry has received attention in \cite{Lebovka1998} for competition interfaces in the Eden growth model, which is in the Kardar-Parisi-Zhang universality class and the competition interface has been shown to be scale invariant with $H=2/3$ see \cite{Saito1995,Grosskinsky2010,Ali2012,Derrida1991}. More recently decreasing domains have also been studied in \cite{lavrentovich2013} under the assumption of diffusive domain boundaries, and our general approach includes both cases. Figure~\ref{bminside}(a) shows an illustration of such a coalescing structure, where particles diffuse with $\gamma=1/2$ on a decreasing radial domain. We can easily adapt the mapping $h(r)$ to take into account the decreasing radius, where (\ref{mapping}) becomes 
\begin{equation}
h(r)= \int_{r}^{r_{0}} \Big(\frac{r_{0}}{s}\Big)^{1/\gamma} ds=\left\{\begin{array}{cl} \frac{\gamma r_{0}}{1-\gamma} \left[ (\frac{r_{0}}{r})^{\frac{1-\gamma}{\gamma}}-1 \right] &,\ \gamma \ne 1  \\ r_{0}\log (\frac{r_{0}}{r}) &,\ \gamma=1 \end{array}\right. , \label{mappingcd}
\end{equation}
where $r_{0}$ is the initial radius. The function (\ref{mappingcd}) is shown in Figure~\ref{bminside}(b), the initial behaviour is $h(r) \simeq r_{0}-r$ and as $r \rightarrow 0$ the limit depends on $\gamma$.

Comparing the mapping for inward growing structures (\ref{mappingcd}) to outward growing structures (\ref{mapping}) we see that (\ref{mappingcd}) has a finite limit for $\gamma>1$, whereas the limit is infinite in (\ref{mapping}), with the opposite behaviour for $\gamma<1$. Although for $\gamma>1$ (\ref{mappingcd}) has a limit
$$h_{\gamma}(\infty)=\frac{\gamma}{\gamma-1} \, r_{0} ,$$
the structure on the fixed domain will have typically already fixated before, since for $\gamma>1$ the fixation time $\tau$ scales as
$$\tau \sim L^{1/\gamma} \ll L \sim r_{0}. $$
So the inward growing structure fixates for all $\gamma>0$. The special case of $\gamma=1$ corresponds to a mirror point where the limit in (\ref{mapping}) and (\ref{mappingcd}) stays the same.

\begin{figure}[t]
\begin{center}
\subfigure[]{\includegraphics[width=3in]{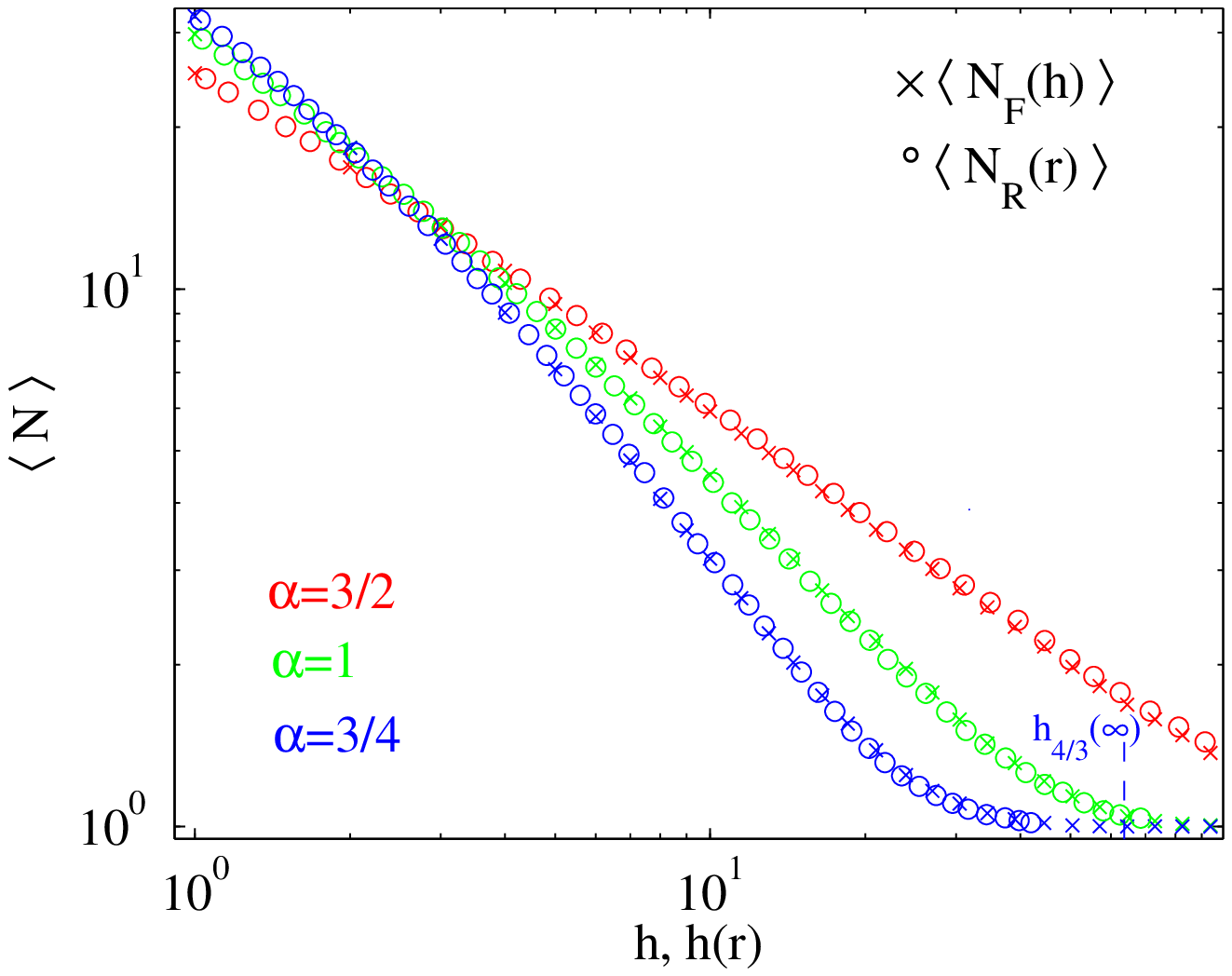}}
\subfigure[]{\includegraphics[width=3in]{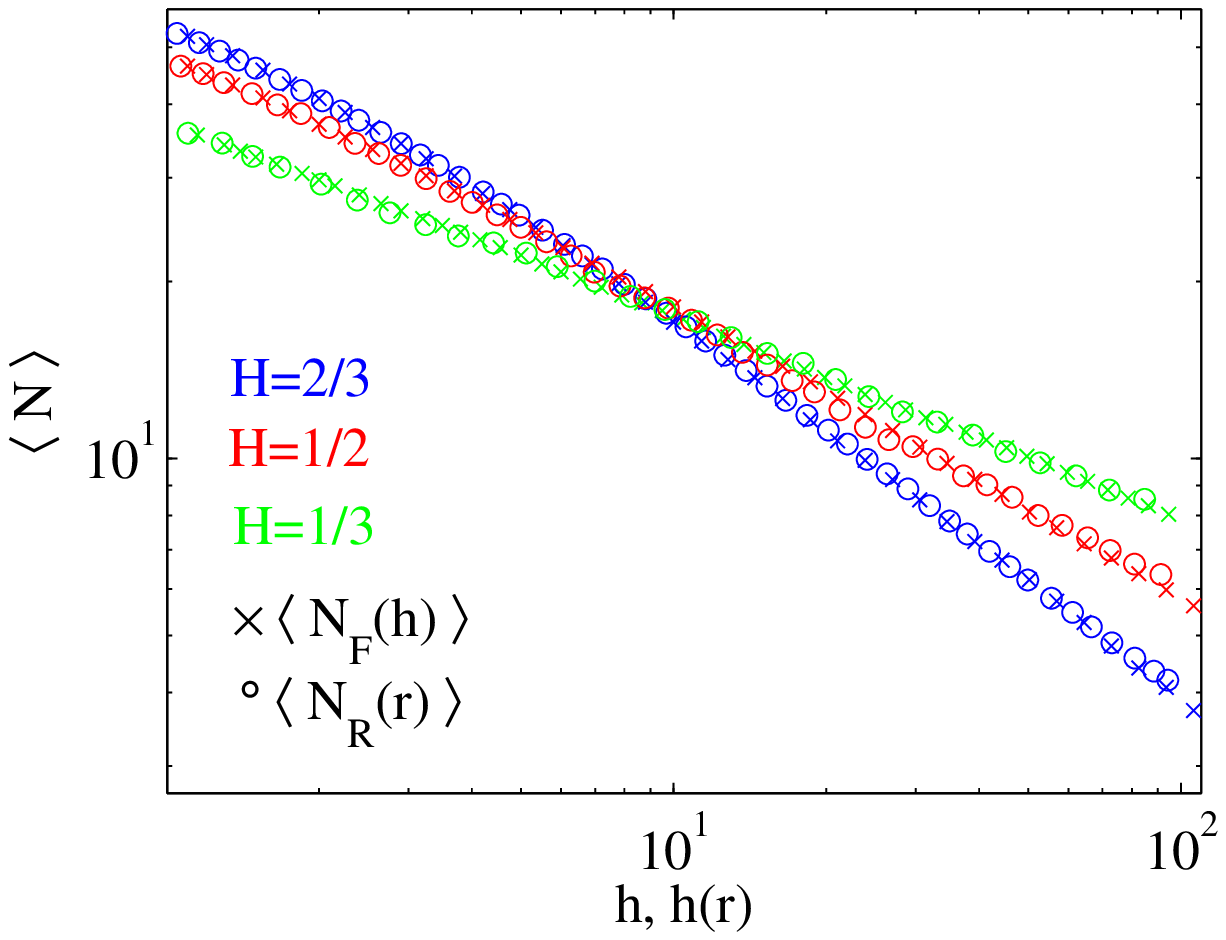}}
\end{center}
\caption{\label{cdwalkers}
The statistics of the fixed ($\times$) and decreasing radial ($\circ$) structures for systems with $L=100$ and $r_{0}=L/2\pi$, respectively, and an initial $100$ arms. We use the inward mapping (\ref{mappingcd}) with $\gamma=\mathrm{max}\{1/\alpha,1/2\}$ for (a) L\'evy structures, and $\gamma=H$ for (b) fBm structures. By plotting $\langle N_{R}(r) \rangle$ vs $h(r)$ we obtain a data collapse. In (a) the limit $h_{4/3}(\infty)\approx 63.66$ corresponding to the data with $\alpha=3/4$ is indicated by a vertical blue dashed line.} 
\end{figure}

In Figure~\ref{cdwalkers} we illustrate the use of (\ref{mappingcd}) for inward growing L\'evy and fBm structures with $\gamma=\mathrm{max}\{1/\alpha,1/2\}$ and $\gamma=H$ respectively. Due to the decreasing size of the domain, $\langle N_{R}(r) \rangle \rightarrow 1$ as $r\rightarrow 0$ and by plotting $\langle N_{R}(r) \rangle$ against $h(r)$ we obtain a data collapse. 
For the value $H=2/3$, results on inward growing radial structures have been seen before in \cite{Lebovka1998}, our approach provides a framework to understand the behaviour in such geometries more clearly. By using the mapping the inward growing behaviour and the outward growing behaviour can be fully described by the fixed domain system.

\subsection{Motion on a general evolving domain}

Consider as before $X:= (X_{h},h\ge 0)$ an arm in the fixed domain, the displacement of this process lies in $[0,L(0))$ for all $h\ge0$. Take $Y:= (Y_{t},t\ge 0)$ to be an arm in a homogeneous time dependent domain, this process will have a displacement in an evolving domain $[0,L(t))$ for all $t\ge0$, where $L(t)$ is a general continuous function such that $L(t) >0$ for all $t\ge 0$. Note that the radially increasing/decreasing domain corresponds to $L(t)=2\pi r(t)$ with $r(t)=r_{0} \pm t$, where in the decreasing case we only consider time $t\in [0,r_{0})$. We assume as before that in each geometry the local scale invariance property (\ref{locsca}) holds. 
The coordinate transformation (\ref{relation}) then generalizes to $x=\frac{L(0)}{L(t)} y$, leading to
$$\frac{dh}{dt}=\Big(\frac{dx}{dy}\Big)^{1/\gamma} = \Big(\frac{L(0)}{L(t)}\Big)^{1/\gamma} .$$
Therefore  \begin{equation} h(t)=\int_{0}^{t}\Big( \frac{L(0)}{L(s)} \Big)^{1/\gamma} ds. \label{mapping1} \end{equation} Analogous to (\ref{mapping}), for $t$ close to $0$ we have $h(t)\simeq t $  and if $L(t) \gg t^{\delta}$ for some $\delta>\gamma$, then we have $$\lim_{t\rightarrow \infty} h(t)<\infty. $$ For Brownian motion and L\'evy flights the rigorous derivation of (\ref{mapping1}) is a simple extension of the approach we have shown in Sections \ref{Bm} and \ref{levymodels}, for fBm a rigorous derivation is given in \ref{asde}.
 
\begin{figure}[t]
\begin{center}
\subfigure[]{\includegraphics[bb=108 277 454 544,clip,height=2.2in,width=2.7in]{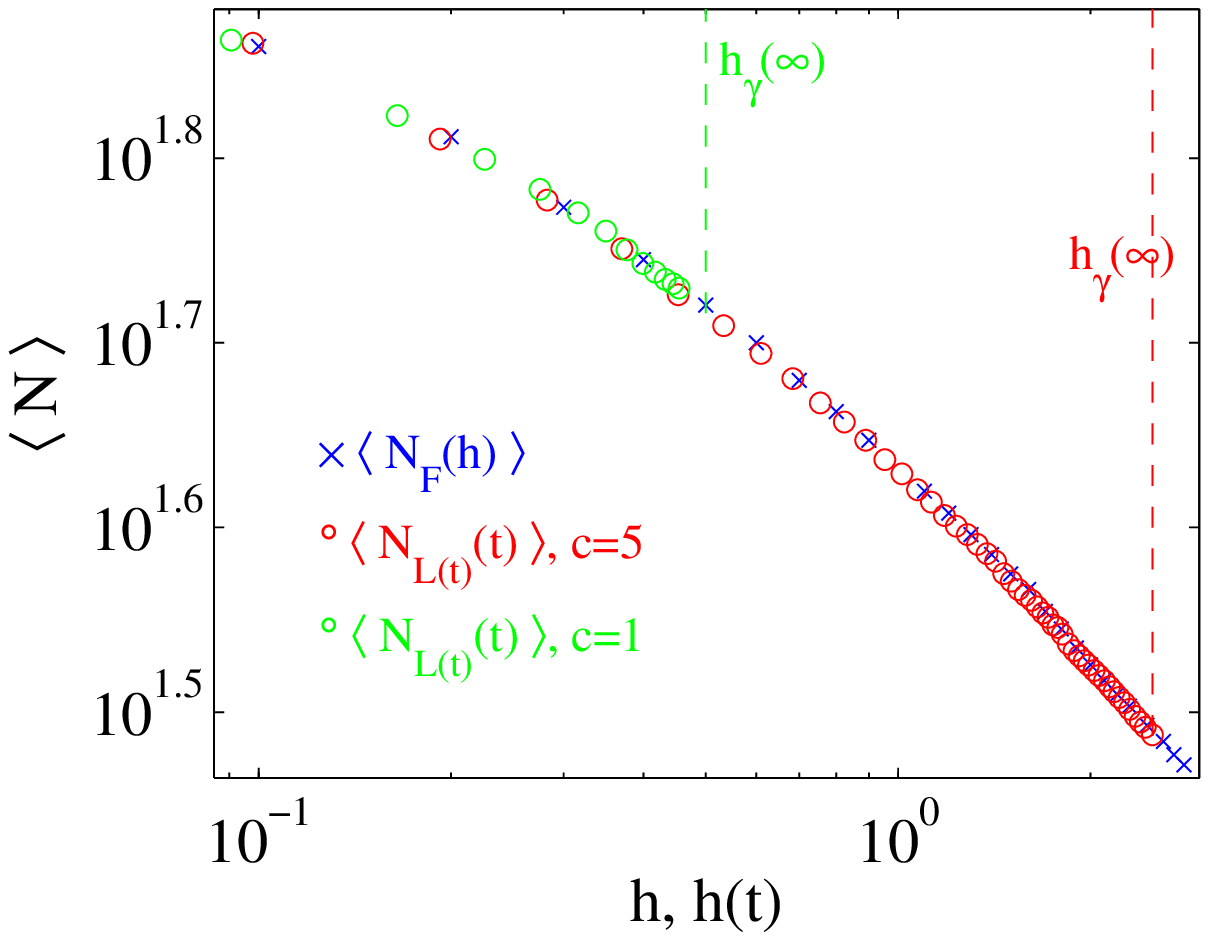}}
\subfigure[]{\includegraphics[bb=109 277 454 544,clip,height=2.2in,width=2.7in]{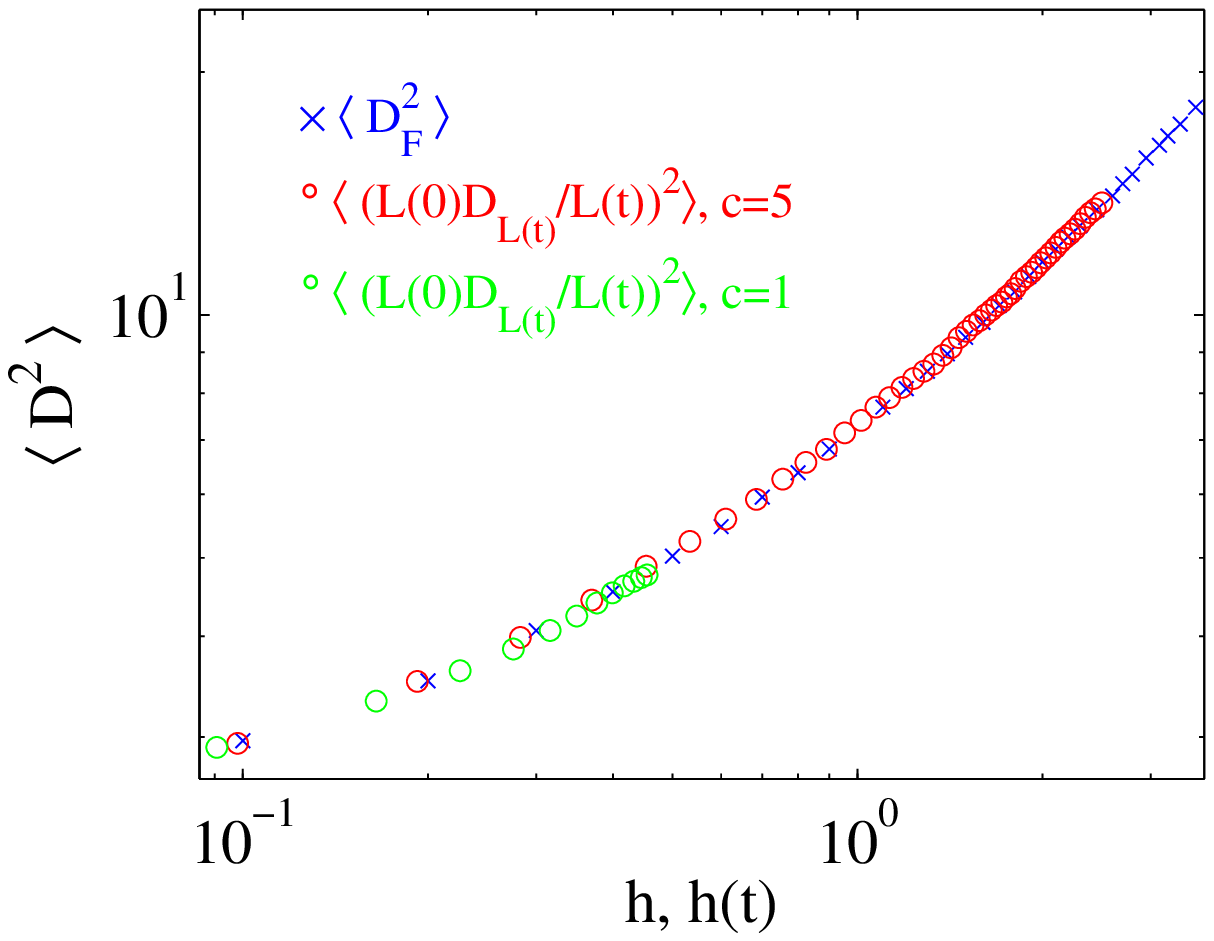}}
\end{center}
\caption{\label{expwalkers}
Illustrating the use of (\ref{mapping1}) for coalescing Brownian motion ($\gamma=1/2$), mapping a growing isotropic structure ($\circ$) to the analogous behaviour in the fixed width structure ($\times$). We choose $L(t)=100 \exp(t/c)$, with $c=1$ and $c=5$ and initially we have $100$ particles. 
By plotting (a) $\langle N_{L(t)}(t) \rangle$ and (b) $\langle(L(0)D_{L(t)}(t)/L(t))^{2}\rangle$ against $h(t)$ we obtain a data collapse. In (a) the color vertical dashed line corresponds to the limit $h_{\gamma}(\infty)=c\gamma$.}
\end{figure}

We use (\ref{mapping1}) to look at the behaviour of a structure evolving in an exponentially increasing domain \be L(t)=L(0)\exp(t/c) \label{function} \ee where $c>0$. Our motivation for choosing such a $L(t)$ is that we can study random walks with an exponentially decreasing jump size which has received a considerable amount of interest (see \cite{kirkpatrick1983,serino2010,krapivsky2004} for more details). These processes have a variety of practical applications in modelling simulated annealing \cite{Torre2000b,Rador2006a} or the displacement of quantum particles \cite{Bressler2007}. For this $L(t)$ we have \begin{equation} h(t)= c\gamma  \left(1-\exp\left(-\frac{t}{c\gamma}\right)\right), \label{mappingexp} \end{equation} and subsequently $h(t)$ has the limit $$h_{\gamma}(\infty)=\lim_{t\rightarrow \infty }h(t)=c\gamma. $$ Note that by the choice of $L(t)$ this limit does not depend on $L(0)$. In Figure~\ref{expwalkers} we illustrate the use of the mapping (\ref{mappingexp}) for coalescing Brownian structures with $\gamma=1/2$. As in previous examples, the behaviour is mapped to the fixed domain by plotting $\langle N_{L(t)}(t)\rangle $ and $\big\langle \big(L(0)D_{L(t)}(t)/L(t)\big)^{2} \big\rangle $ against $h(t)$.

\subsection{Structures in $n+1$ dimensions}
It is interesting to note that the general mapping (\ref{mapping1}) does not depend on the dimensions $n$ of the state space. Consider a $n+1$ dimensional structure, where in each spatial direction the displacements are $Y_i\in [0,L_i (t))$ with $L_i (t)>0$ for all $t \ge 0$. We can still characterize the behaviour of this $n+1$ dimensional evolving structure by mapping it to a fixed width structure where in each spatial direction the displacements $X_i$ are in the fixed domain $[0,L_i (0))$. In order to do this, we require the local scale invariance property (\ref{locsca}) to hold in all spatial directions $i=1,\ldots ,n$
\be\label{dxi}
  dX_i \sim (dh)^\gamma \quad\mbox{and}\quad d Y_i\sim (dt)^\gamma \,.
\ee
For such systems the mapping (\ref{mapping1}) stays exactly the same and will therefore map an evolving structure to a fixed width structure. It is also possible to include anisotropy in (\ref{dxi}) where there can be a possible $i$-dependence of the multiplicative factors 
but $\gamma$ must be identical in all directions. This is illustrated in Figure~\ref{lengthwalkers}(b) in Section \ref{non-local} for coalescing particles of non-zero size on a growing 2-sphere.

\section{Generalized local interactions \label{non-local}}
In this section we extend our theory to systems with non-local interactions. As relevant examples we consider coagulating structures where particles either have a non-zero size $d>0$, or structures with particles that coagulate and branch. We illustrate the mapping (\ref{mapping}) for radially increasing and fixed domain structures, which are composed of particles that perform Brownian motion with $\gamma=1/2$. 

\subsection{Particles with non-zero size}

\begin{figure}[t]
\begin{center}
\subfigure[]{\label{lennum}\includegraphics[width=3in]{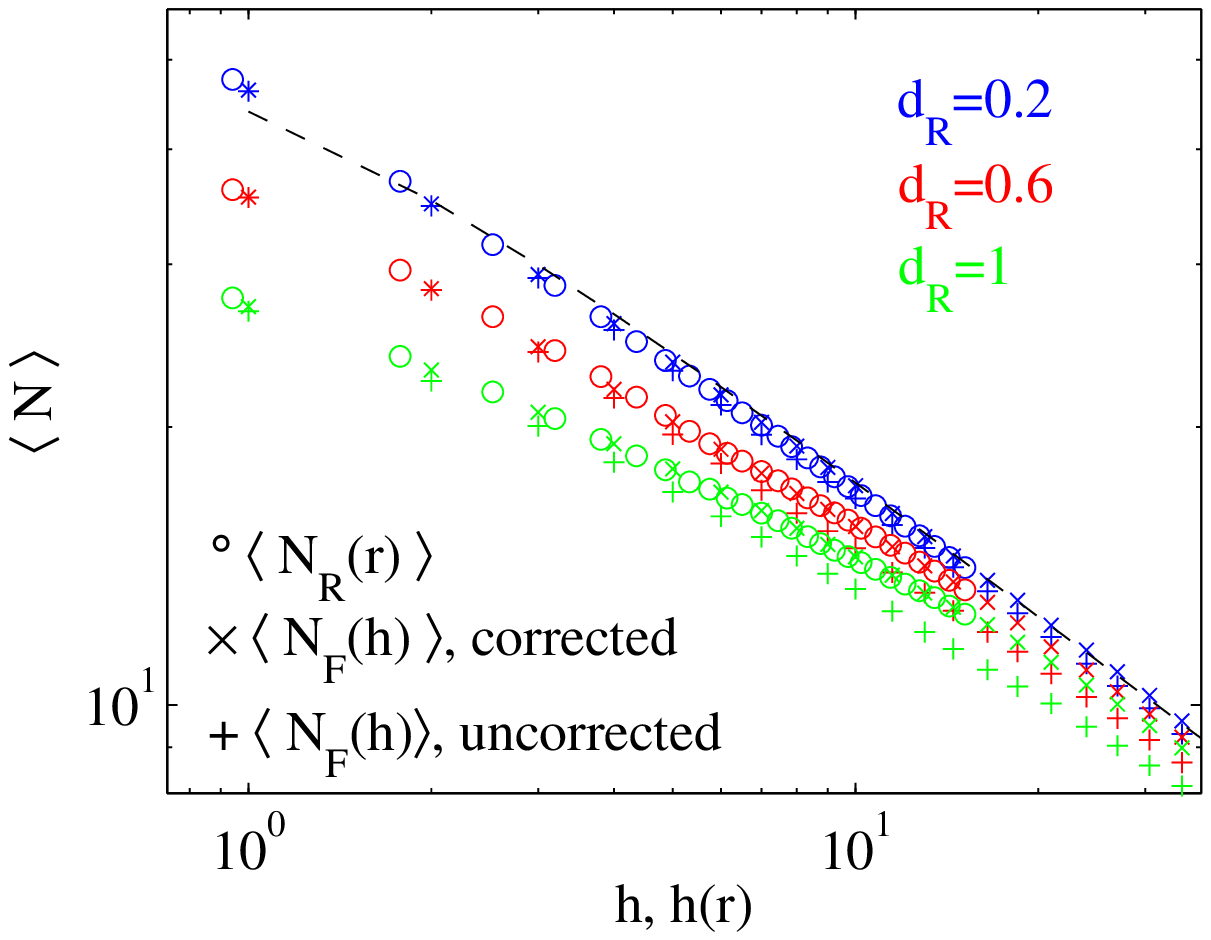}}
\subfigure[]{\label{lendist}\includegraphics[width=3in]{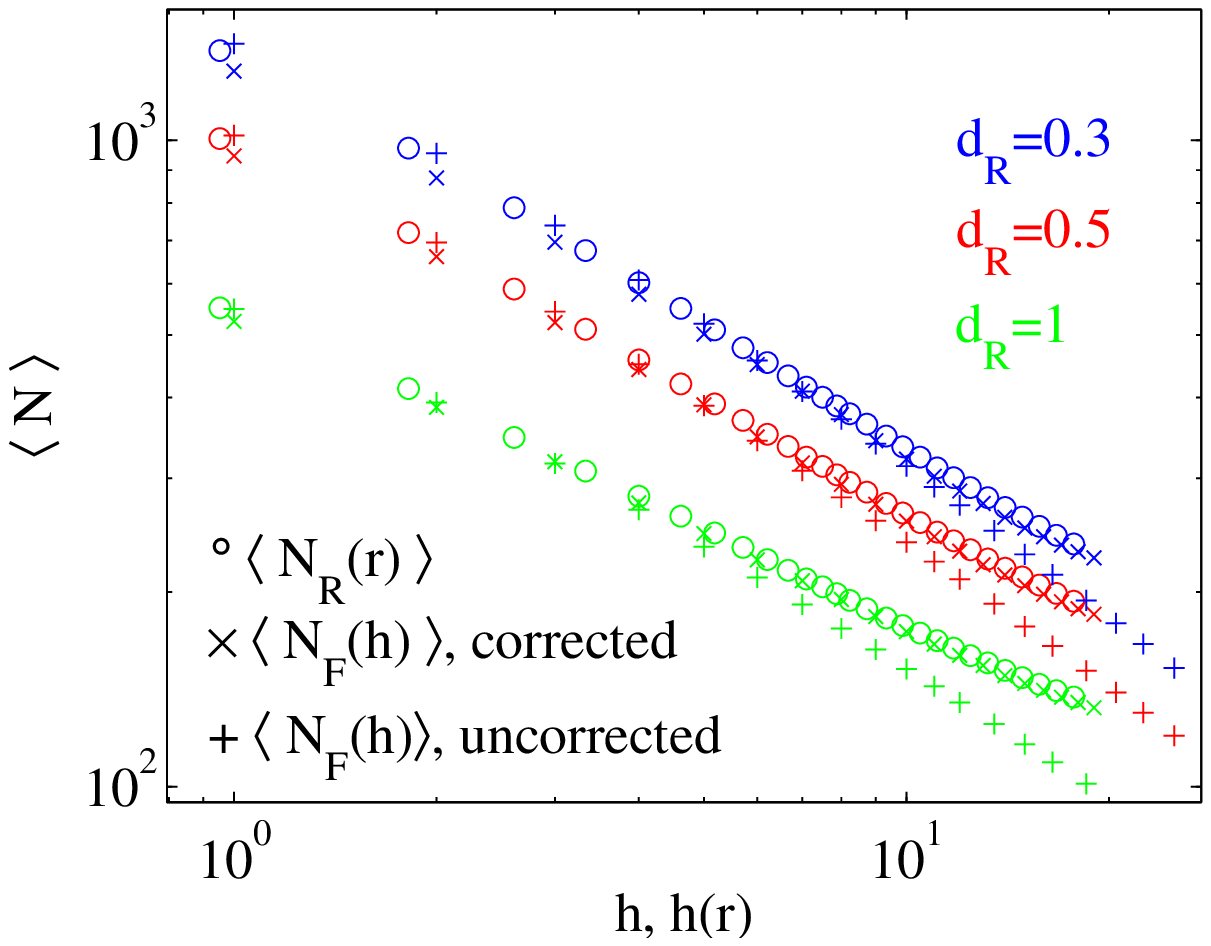}}
\end{center}
\caption{\label{lengthwalkers}
Mapping the radially increasing structure to the fixed width structure for a system with non-local interactions. Here particles perform Brownian motion, they have a diameter $d>0$ and coalesce when the distance between their centers is less then $d$. We use (\ref{mapping}) with $\gamma=1/2$ and initially we have $100$ particles. (a) Mapping $\langle N_{R}(r) \rangle $ in $1+1$ dimension with $r_{0}=100/(2\pi )$. Dashed black line is the analytical prediction (\ref{lsec}), as seen in figure~\ref{fbmstats}(a). (b) Mapping $\langle N_{R}(r) \rangle $ in $2+1$ dimension with $r_{0}=20$.  
In each case the relationship (\ref{diam}) is used to obtain an exact mapping ($\times$), compared to an approximate mapping ($+$) when (\ref{diam}) is not used. }
\end{figure}

Most real world structures exhibit a microscopic length scale, which in our representation corresponds to a non-zero particle size $d>0$; this influences the structures on small length scales \cite{Kopelman1988}. Here we look at coalescing particle systems, where each particle has an isotropic shape with a diameter $d>0$. Introducing such a length scale in the interactions means that particles will now coagulate when the distance between their centers is less then $d$. As long as this is much smaller than the system size, i.e. $d \ll L$, the corrections introduced are small (see Figure~\ref{lengthwalkers}). We include such corrections into the mapping by preserving the particle size scale in each domain relative to system size. Taking $d_{R}$ as the fixed diameter in the radial geometry, we have
\be d_{F} =\frac{r_{0}}{r(h)}d_{R} \label{diam}, \ee
where $d_{F}$ is the rescaled diameter in fixed geometry such that
$$d_{F} \rightarrow 0 \quad \mbox{as} \quad h\rightarrow h_\gamma (\infty ) .$$
The function $r(h)$ in (\ref{diam}) is the inverse of (\ref{mapping}) and for general $\gamma \ne1 $ it has the form
\begin{equation} r(h)=r_0 \left( 1-\frac{1-\gamma}{\gamma}\,\frac{h}{r_0}\right)^{-\gamma /(1-\gamma)}
\label{invmap} \end{equation}
In Figure~\ref{lengthwalkers}(a) we look at such systems for a range of diameters $d$. By using the mapping (\ref{mapping}) with $\gamma=1/2$ we are able to map the behaviour in the radially growing structure to the fixed width structure, illustrated for $\langle N_{R}(r) \rangle $. For the fixed width structure simulations we include the data where the correction (\ref{diam}) is applied ($\times$) and where it is ignored ($+$). We can see that the inclusion of (\ref{diam}) provides an exact mapping between the two domains.
The introduction of a particle size only affects the initial behaviour, where initially the distance between particles is small and due to $d>0$ more coalescing events take place. As time increases, the distance between particles increases and the behaviour becomes largely independent of $d$.

In Figure~\ref{lengthwalkers}(b) we show that the mapping also works in $n=2$ dimensions, where using (\ref{mapping}) we map the data of $\langle N \rangle$ from a growing sphere $S^{2}(r)$ to a fixed sphere $S^{2}(r_{0})$. As above, particles have a given diameter $d>0$ and we use the correction (\ref{diam}) on the fixed width structure as indicated by ($\times$) to obtain an exact data collapse as oppose to an approximation, indicated by ($+$) when (\ref{diam}) is not used. Note that in $2$ dimensions particles are not ordered, so in order to detect all coagulation events correctly in the simulation we have to choose increments in the particle motion small enough compared to the diameter $d$.

\subsection{Structures with branching}
A similar treatment is possible for more general interactions with intrinsic time scales. Here we treat coagulating and branching structures. These structures have much interest due to a wide variety of applications, some examples include the modelling of surnames in genealogy \cite{Popovic2005}, or the growth of microbial species with mutation \cite{lavrentovich2013}. We generalize the diffusing coalescing model studied in Section~\ref{secresults} by adding the mechanism of particle branching to the system. In order to map the length scale of branching correctly, which is encoded in the branching rate, 
we derive a relationship between the rate $R_{F}$ in the fixed domain and $R_{R}$ in the growing radial domain such that the number of branch events in each domain is equal. Let $N_{R}(\Delta r)$ be the number of branch events in the radial domain in the interval $[r,r+\Delta r]$ and let $N_{F}(\Delta h)$ correspond to the number of events in the fixed domain in the interval $[h,h+\Delta h]$. Then $R_{F}=N_{F}(\Delta h)/\Delta h$ and $R_{R}=N_{R}(\Delta r)/\Delta r$, and by using (\ref{mapping}) and requiring $N_{F}(\Delta h)=N_{R}(\Delta r)$, we have $$\frac{R_{R}}{R_{F}}= \frac{N_{R}(\Delta r)/\Delta r}{N_{F}(\Delta h)/\Delta h}= \frac{\Delta h}{\Delta r} =\frac{r_{0}^{2}}{r^{2}}. $$ 
Mostly we are interested in mapping a radial structure with fixed rate $R_{R}$ to a fixed width structure with a variable rate such that
\begin{equation} R_{F}(h)=\Big(\frac{r(h)}{r_{0}}\Big)^{2}  R_{R}, \label{rates2}\end{equation} 
where $r(h)$ is as in (\ref{invmap}) with $\gamma=1/2$. 
 
We consider two types of models where particles branch. In each case the particles perform Brownian motion and coalesce upon contact, and a branch event occurs after a random time exponentially distributed with mean $1/R_{F}$ ($1/R_{R}$) for the fixed (radial) domain structure. In one model, which we call ``uniform'', the new particle is placed uniformly in the domain. In the second model which we call ``local'' the new particle is placed in the same position as its mother. In order to avoid instantaneous coalescence between the daughter and mother particles we prohibit coalescence and both particles move independently until they have both branched again.

\begin{figure}[t]
\begin{center}
\subfigure[]{\includegraphics[width=3in]{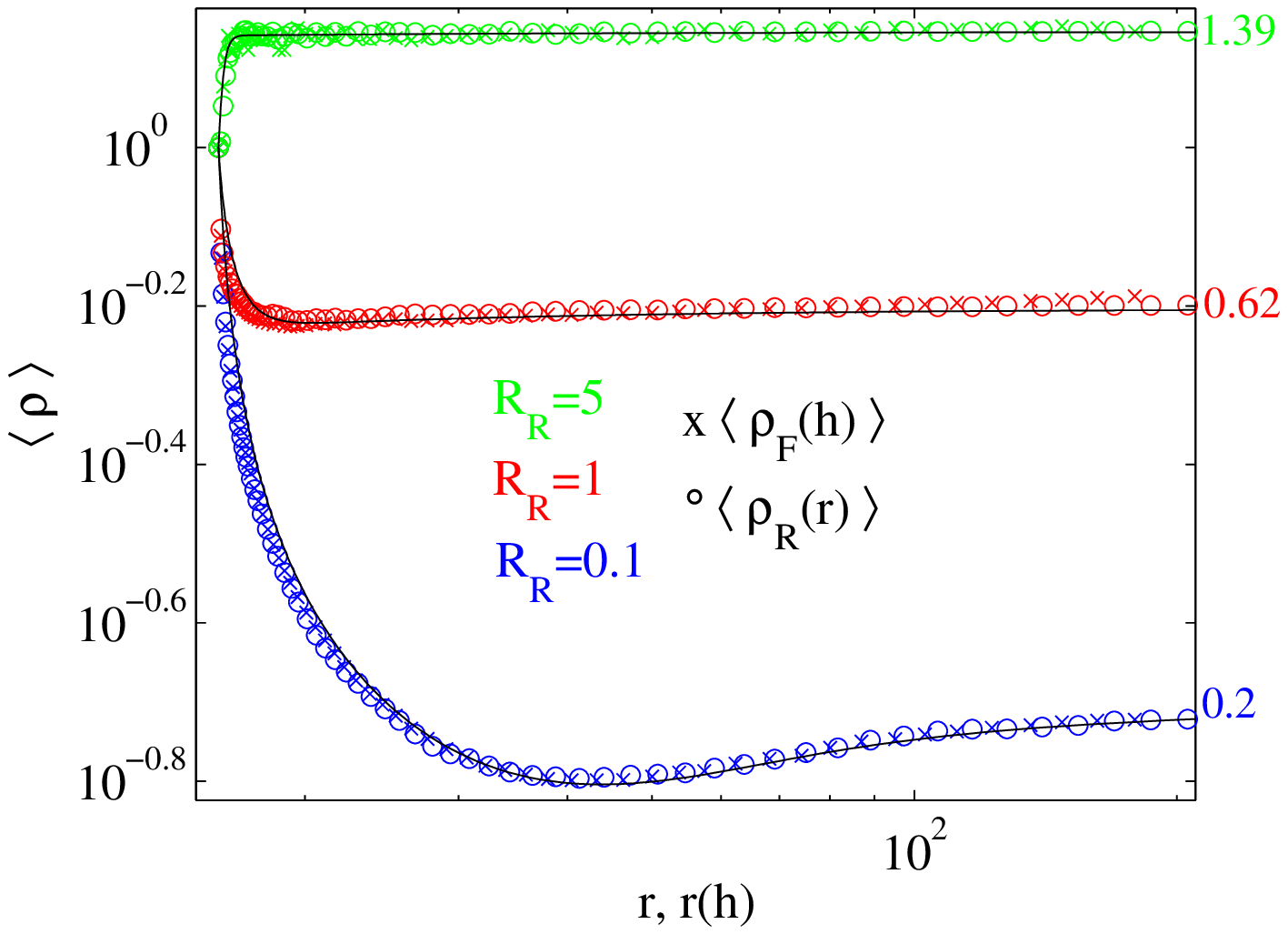}}
\subfigure[]{\includegraphics[width=3in]{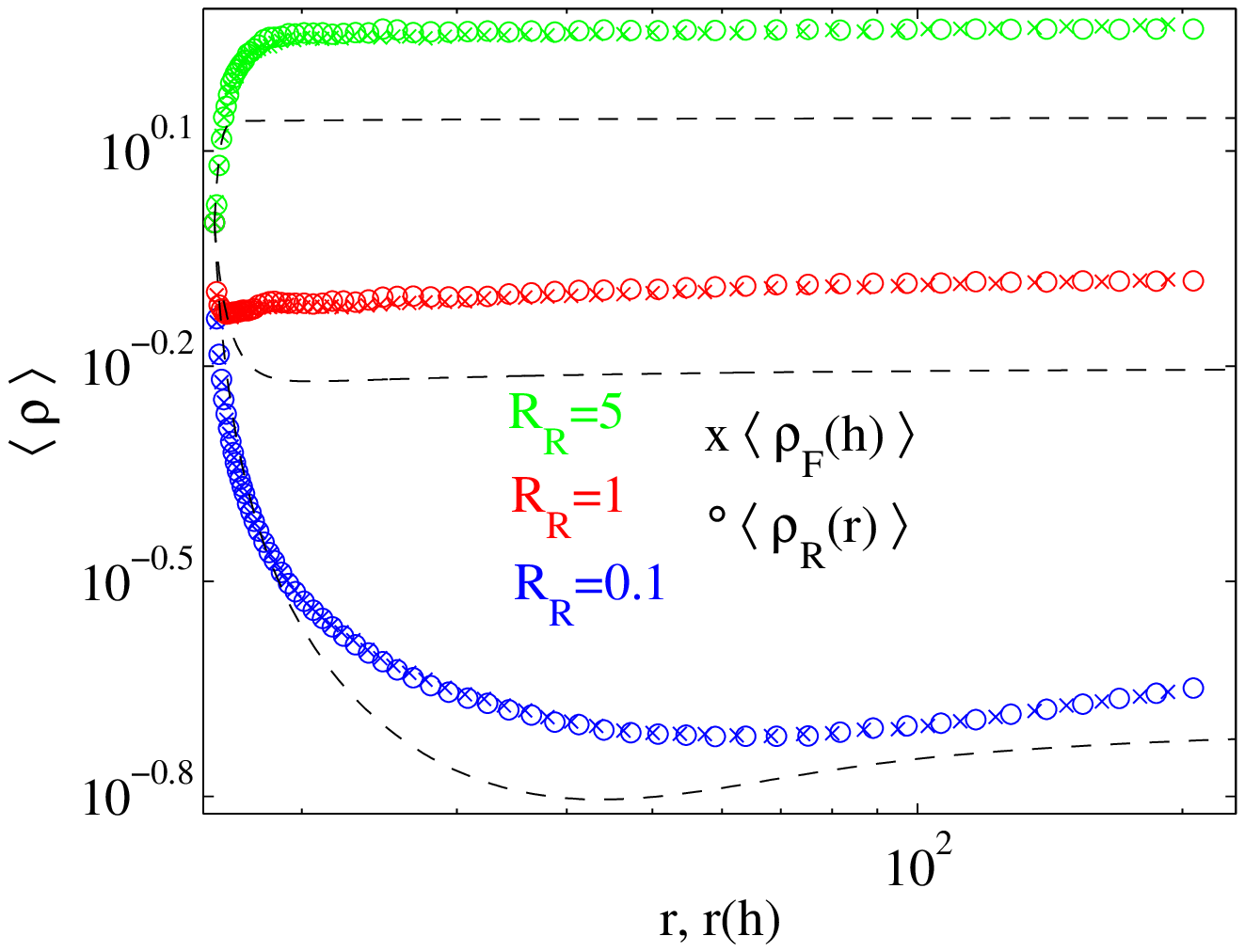}}
\end{center}
\caption{\label{bmbranchdivedbyr}
Mapping the fixed domain to the growing radial domain for the branching models (see text). We use (\ref{invmap}) with the relationship (\ref{rates2}), simulations are performed for systems with $r_{0}=L/(2\pi)$ and $L=100$ with $100$ initial particles.  We use several values for $R_{R}$ and in each case plot $\langle \rho_{F}(h) \rangle $ vs $r(h)$ for (a) the uniform model, and (b) the local model and obtain a data collapse. The solid black line in (a) is the analytical prediction (\ref{denbranchuni2}) with limits (\ref{denbranchuni3lim}) indicted on the the right axis. 
Since this prediction is based on mean-field agruments, it does not work well for the local model, as is shown in (b) (black dashed line) for comparison.}
\end{figure}

In Figure~\ref{bmbranchdivedbyr} we use the mapping (\ref{invmap}) with the relationship (\ref{rates2}) to map the density from the fixed domain to the growing radial domain. This shows that $\langle N_{R}(r) \rangle$ is asymptotically linear in $r$ and we can predict the speed constant. In (a) the full black line and in (b) the dashed black line correspond to the expression (\ref{denbranchuni2}) which is an analytical prediction for the density $\rho(r)$ for the uniform model. Here we have adapted previous mean field results, \cite{Lebovka1998,Alemany1995,Avraham2001}, to take into account the density dependent input of particles, as explained below.

For coalescing structures in the fixed domain, when particles perform Brownian motion the one dimensional rate equation \cite{Lebovka1998} governs the behaviour of the density $\rho(h)$, and reads \begin{equation} \frac{d \rho}{dh}=-\frac{\pi \rho^{3}}{2} \label{mflin} \end{equation} with solution $\rho(h) \sim h^{-1/2}.$
For the uniform model we adapt Eq.~(\ref{mflin}) by considering a density dependent input into the domain \cite{Avraham2001}. This leads to an additive term
\begin{equation} \frac{d \rho}{dh}=-k_{1}\rho^{3}+k_{2}\rho \label{mflinbr}, \end{equation}
where $k_{1}=\pi/2$ and $k_{2} =C_2 R_F$ is a model dependent constant proportional to $R_{F}$. Fitting to the data we find $C_{2}\approx 0.6158$. 
In order to obtain an analytical expression for the growing radial domain, we modify the density equation (\ref{mflinbr}) using the relation (\ref{rates2}), which leads to
\begin{equation} \frac{d \rho}{dh}=-k_{1}\rho^{3}+k_{2}\Big(\frac{r_{0}}{r_{0}-h}\Big)^{2}\rho \label{mflinbrnew}. \end{equation}
Using (\ref{mapping}) the solution to (\ref{mflinbrnew}) can be written as 
\be \rho(r) = \frac{e^{k_{2} r}r_{0}}{r} \Bigg[e^{2k_{2} r_{0}}+2k_{1}  \Big[-e^{2k_{2} r_{0}} r_{0}^{2}/t+2k_{2} r_{0}^2 \Phi\big(2k_{2} t\big) \Big]_{r_{0}}^{r} \Bigg]^{-1/2} \label{denbranchuni2}, \ee
where $\Phi(x)=\int^{x} e^{t}/t \,dt$. For large $r$ (\ref{denbranchuni2}) can be expressed as \be \rho(r) \approx \frac{1}{r} \Bigg[2k_{1}  \Big(-\frac{1}{r}+2k_{2} e^{-2k_{2} r_{0}}\Phi\big(2k_{2} r\big) \Big) \Bigg]^{-1/2} \label{denbranchuni3}, \ee
and for $r\rightarrow \infty$ $$e^{-2k_{2} r_{0}}\Phi\big(2k_{2} r\big) \approx \frac{1}{2k_{2} r}+ \frac{1}{(2k_{2} r)^{2}} +O(1/r^{3}). $$
Combining this with (\ref{denbranchuni3}) gives 
\be \lim_{r\rightarrow \infty } \rho(r) = \left(\frac{k_{2}}{k_{1}}\right)^{1/2}\approx \sqrt{\frac{C_2  R_{R}}{\pi/2}} \label{denbranchuni3lim}. \ee
This limit is indicated on the right axis in Figure~\ref{bmbranchdivedbyr}(a) and matches the data very well, along with the prediction of the full solution (\ref{denbranchuni2}).

The validity of the rate equation is based on mean-field arguments, and it is therefore not surprising that it does not predict well the behaviour of the local model, as can be seen in Figure~\ref{bmbranchdivedbyr}(b). However, the main point is that we still have a perfect data collapse from the mapping and can use the fixed geometry to fully understand the radially increasing domain. Other processes such as pair creation, which are relevant in mutation processes in expanding biological populations \cite{lavrentovich2013}, can be treated in full analogy.

\section{Discussion}
In this paper we have studied behaviour of two and three dimensional growth structures which can be described by trajectories of interacting particles in time dependent domains. For Markovian processes, which are completely determined by their local scale invariance properties, we could derive a general mapping, which describes time-dependent domains in terms of a non-linear time change. This description is universal in the sense that it has the scaling exponent as the only parameter, and is independent of other factors such as dimensionality. Exact calculations for fractional Brownian motion, a scale invariant process with memory, lead to a weaker result on moment matching with a corrected version of the mapping, which is numerically almost identical to the general one. Our approach also covers various local particle interactions and can be adapted to non-local ones, and is therefore applicable in a wide range of applications. It can be used to effectively study spatial competition dynamics, such as the interfaces in a Eden growth model \cite{Saito1995} with biological applications as seen in \cite{Hallatschek2007,Ali2012}. Our results can also be applied to various other physical systems, such as understanding the motion of advected particles in turbulent fluids \cite{Bohr1993} or simulated annealing \cite{krapivsky2004,Rador2006b}.
There are of course various growth models which do not immediately fall in the class of systems we treated, such as diffusion limited aggregation (DLA) or viscous fingering where the local growth rates depend on the entire geometry of the cluster \cite{Witten1981,Saffman1958,Amar1991}. However, we have promising numerical evidence that DLA can be effectively treated by a similar mapping approach, which is current work in progress. Another interesting extension of our results would be to see whether long range interactions or non-neutral particle evolution with drift can be included in our approach to describe even more general situations.

\section*{Acknowledgments}
This work was supported by the Engineering and Physical Sciences Research Council (EPSRC), Grant No. EP/E501311/1.

\appendix

\section{Analytical form for $\langle N_{F} \rangle $ and $\langle D^{2}_{F} \rangle $}

In this section we derive expressions for the analytical behaviour of the statistics as shown in Figure~\ref{levystats} and Figure~\ref{fbmstats}. For the fBm structures the behaviour is an approximation of exact results for Bm, whereas for L\'evy flights we adapt previously known mean field results.

\subsection{Fractional Brownian Motion (fBm) \label{ssA1}}

In this section, we use the method of empty intervals to find the inter-particle distribution function (Ipdf) $E(x,t)$, which can be used to predict the number of particles $\langle N_{F}(t) \rangle $ for walkers that undergo coagulation.  
For a full review of the theory see \cite{Alemany1995,Sasaki2000}.
Define a function $E(x,t)$ to be the probability that an arbitrary interval of length larger or equal to $x$ is empty at time $t$. The concentration of particles $\rho(t)$ is then given by
\begin{equation}\label{rela1}
\rho(t)=-\frac{\partial E}{\partial x} \Big\vert _{x=0}.
\end{equation}
For a system with finite size say $L$ we define the function $E: [0,L] \times \mathbb{R}_{+} \rightarrow [0,1]$. The method of empty intervals relies on the Chapman-Kolmogorov equation of the process and for fBm, $E(x,t)$ will satisfy the following partial differential equation (pde) \cite{Hahn2011}
\begin{equation} \frac{\partial E}{\partial t}=2\sigma^{2}Ht^{2H-1}\frac{\partial^{2} E}{\partial x^{2} } ,\label{pde1} \end{equation}
$\sigma^{2}$ is the prefactor of the mean squared displacement $\langle X_{t}^{2} \rangle$. The relation (\ref{rela1}) further requires that the history before a coagulation event is irrelevant for the future time evolution. This is not the case for fBm, so the following calculation is not exact, but turns out to give a good approximation.

The solution of (\ref{pde1}) should satisfy the Dirichlet boundary conditions $$E(0,t)=1 \quad \mbox{and} \quad E(L,t)=0 \quad \mbox{for all} \quad t\ge0 $$ and if particles initially have a fixed distance of $1$ then the initial condition $$E(x,0)=1_{x \le 1 } \quad \mbox{for all} \quad x\in [0,L)$$ holds. In order to solve (\ref{pde1}) we consider the transformed equation \begin{equation} \frac{\partial E}{\partial T}=\sigma^{2}\frac{\partial^{2} E}{\partial x^{2} } \label{pde2} \end{equation} 
obtained by taking $T=t^{2H}$. To solve (\ref{pde2}), we construct a free-space Greens function, $V(x,T)$, which is a solution to the following equation \begin{equation}-\frac{\partial V}{\partial T} -\sigma^{2}\frac{\partial^{2} V}{\partial x^{2} }=\delta(x-x')\delta(T-T'). \label{gpde}\end{equation}
We can show the solution of (\ref{gpde}) to be 
\begin{equation}
V(x,T,x',T')=\left\{\begin{array}{cl} \frac{1}{\sqrt{(4\sigma^{2}(T'-T)\pi)}}\exp(-\frac{(x-x')^{2}}{4\sigma^{2}(T'-T)}) &,\ T\le T' \\ 0&,\ T> T' \end{array}\right. . \label{Vsol}
\end{equation}

Using the free-space Greens function $V(x,T,x',T')$ we construct a Greens function $G(x,T,x',T')$ that satisfies the following relation

\begin{equation} \int_{0}^{L} \Big[EG \Big]^{\infty}_{0} dx+\sigma^{2}\int_{0}^{T'} \Big [E\frac{\partial G}{\partial x}- G\frac{\partial E}{\partial x}\Big ]^{L}_{0} dT =-E(x',T'). \label{solvingpdeG}\end{equation}
The Greens function has to satisfy the boundary condition $G(0,T,x',T')=0$ and $G(L,T,x',T')=0$. This leads us to the following form of the Greens function 
\begin{equation} G(x,T,x',T')=\!\! \sum_{n=-\infty}^{\infty} [V(x{-}2nL,T,x',T')-V(x{-}2nL,T,-x',T')] \end{equation} and $G(x,T,x',T')$ satisfies the pde (\ref{gpde}).

Evaluating (\ref{solvingpdeG}), we have $$E(x',T')=\int_{0}^{1}G(x,0,x',T')dx+\sigma^{2} \int_{0}^{T'} \frac{\partial G}{\partial x} \Big\vert_{x=0} dT. $$ Further evaluation of the last integral leads to 
$$\frac{2}{\sqrt{\pi}} \sum_{n=-\infty}^{\infty} \int_{\frac{|x+2nL|}{\sqrt{4\sigma^{2}T}}}^{\infty}\exp(-u^{2})\,du.$$ 
and with (\ref{rela1}) we get
\begin{eqnarray} \rho(T) &=& \frac{1}{\sqrt{\pi \sigma^{2} T}}\sum_{n=-\infty}^{\infty}\Big[\exp\Big(-\frac{(1-2nL)^{2}}{4 \sigma^{2}T}\Big) - \exp\Big(-\frac{L^{2}n^{2}}{\sigma^{2}T}\Big)\Big]\nonumber\\
& &\ +\frac{\vartheta_{3}\Big(0,e^{-\frac{L^{2}}{\sigma^{2} T}}\Big)}{\sqrt{\pi \sigma^{2} T}}\ .\label{roform}
\end{eqnarray}
Here $\vartheta_{3}(\cdot,\cdot)$ is the elliptic theta function of the third kind, defined as  $$\vartheta_{3}(z,q)=1+2\sum_{n=1}^{\infty } q^{n^{2}}\cos (2n z),$$ where $z,q\in \mathbb{C}$ and $|q|<1$.

Using $T=t^{2H}$ the average number of particles $\langle N_{F}(t) \rangle$ takes the form
\begin{eqnarray}
\langle N_{F}(t) \rangle &= L \Bigg[ \frac{1}{\sqrt{\pi \sigma^{2} t^{2H}}}\sum_{n=-\infty}^{\infty}\Big[\exp\Big(-\frac{(1-2nL)^{2}}{4 \sigma^{2}t^{2H}}\Big) - \nonumber \\ &\qquad\quad \exp\Big(-\frac{L^{2}n^{2}}{\sigma^{2}t^{2H}}\Big)\Big]+\frac{\vartheta_{3}\Big(0,e^{-\frac{L^{2}}{\sigma^{2} t^{2H}}}\Big)}{\sqrt{\pi \sigma^{2} t^{2H}}} \Bigg], \label{lsec}
\end{eqnarray}
where we have used $\langle N_{F}(t) \rangle=L\rho(t^{2H})$.

In order to calculate an analytical prediction for $\langle D_{F}^{2} \rangle $, we use the inter-particle distance pdf $p(x,t)$, which can be computed as \begin{equation} p(x,t)=\rho(t)^{-1} \frac{\partial ^{2} E}{\partial x^{2}}. \label{pdfdist} \end{equation}

Using (\ref{pdfdist}) we have
$$\langle D_{F}^{2} \rangle = -2\rho(t)^{-1} \int_{0}^{L} x\frac{\partial  E}{\partial x}
dx, $$
this leads to 
\begin{eqnarray} \lefteqn{\langle D_{F}^{2} (t)\rangle  =-2\rho(t)^{-1} \int_{0}^{L} dx \Bigg [ \frac{1}{\sqrt{4\pi t^{2H} \sigma^{2}}} \sum_{n=-\infty}^{\infty} \Big[ 2\exp\Big(-\frac{(x-2nL)^{2}}{4t^{2H}\sigma^{2}}\Big) -} \nonumber \\
& &\exp\Big( {-}\frac{(x{-}2nL{+}1)^{2}}{4t^{2H}\sigma^{2}}\Big) {-}\exp\Big( {-}\frac{(x{+}2nL{-}1)^{2}}{4t^{2H}\sigma^{2}}\Big) \Big] {-} \frac{\vartheta_{3}\Big(\frac{\pi x}{2L},e^{-\frac{-\pi^{2} \sigma^{2}t^{2H}}{L^{2}}} \Big)}{L} \Bigg] . \label{fbmdistspred} \end{eqnarray}

\subsection{L\'evy Flights \label{ss2A1}}

Attempting to compute the prediction $\langle N_{F}(t) \rangle $ for coalescing L\'evy structures in the fixed domain with the empty interval method leads to a pde with fractional derivatives. This pde is not well posed so no solution in closed form can be derived. We thus use the general rate equation from \cite{Lebovka1998,Alemany1995,Avraham1990}, which governs the long time dynamics of the density $\rho(t)$. The form of the rate equation follows from (\ref{roform}), where as $T=t^{2\gamma }\rightarrow \infty$ and for large system sizes (i.e. $L\rightarrow \infty$) we have $$\rho(t)=\frac{1}{\sqrt{\pi \sigma^{2} t^{2\gamma}}}.$$
Using $\gamma =1/\alpha$ and the fact that $\sigma =\sigma_\alpha$ depends on $\alpha$ as in (\ref{levypdf}) we get for the number of particles
\be \langle N_{F}(t) \rangle  =\frac{L}{\sqrt{\pi \sigma_{\alpha} t^{2/\alpha}}}. \label{lnumpred} \ee

The form of $\langle D_{F}^{2} (t) \rangle $ can be derived as follows. From (\ref{fbmdistspred}) the long time and large scale (i.e. $L \rightarrow \infty$)
behaviour for fBm structures is $$\langle D_{F}^{2} (t)\rangle = 4 \sigma^{2} t^{2\gamma},$$ where $\gamma=H$. For the analogous fixed L\'evy structures using that $\gamma=1/\alpha$, we have \be \langle D_{F}^{2} (t)\rangle \sim t^{2/\alpha} .\label{ldistsqpred}\ee
The power has been confirmed in Figure~\ref{levystats}(b) with fitted prefactors.

\section{Deriving the mapping for fBm with fractional calculus\label{asde}}
When the arms of the structures are fBm $B^{H}=(B^{H}_{t}, t\ge 0)$ with $H \in (0,1)$ and on a general evolving domain $[0,L(t))$, with $L(t)$ continuous and $L(t)>0$ for all $t\ge0$, we can derive the generalized form of the mapping (\ref{mappingfbm1}) using It{\^o} isometry. Consider the rescaled process $dZ_t =\frac{ L(0)}{L(t)}\,dB^{H}_{t}$ in integral form 

\be \label{Zfbm1}
  Z_t = \int_{0}^{t}\frac{ L(0)}{L(s)} \,dB^{H}_{s},
\ee
where $B^{H}_{t}$ is a standard fBm process. 
For $H \ne 1/2$ in order to treat (\ref{Zfbm1}) with It{\^o}-calculus, we use an isometric memory kernel $K_{H}$, where details are given in \cite{Biagini2008}. 
We can represent (\ref{Zfbm1}) as
\begin{equation}
Z_{t}= \int_{0}^{t} \Big(K_{H}*\frac{L(0)}{L(\cdot)} \Big)(s)dB_{s} \label{sdefbm1}, 
\end{equation}
where $B_{t}$ is a standard Brownian motion. 
The form of the operator is  $$\Big (K_{H}*\frac{L(0)}{L(\cdot)}\Big)(s)=\int_{s}^{t}\frac{L(0)}{L(s')}\frac{\partial k_{H}(s',s)}{\partial s'}ds' $$
where 
$$\frac{\partial k_{H}(s',s)}{\partial s'}= c_{H}\Big(\frac{s'}{s}\Big)^{H-1/2}(s'-s)^{H-3/2} $$ and $$c_{H}=\Big(\frac{(H(2H-1))}{\beta(2-2H,H-1/2)}\Big)^{1/2}\quad \mbox{and} \quad\beta(a,b)= \frac{\gamma(a+b)}{\gamma(a)\gamma(b)}.$$

Define
\be f(t,s)= \frac{c_{H} L(0)}{s^{H-1/2}} \int_{s}^{t}\frac{s'^{H-1/2}(s'-s)^{H-3/2}}{L(s')} \, ds' \nonumber
\ee
such that
\begin{equation} Z_{t}=\int_{0}^{t} f(t,s) dB_{s} \label{sdefbm2}. \end{equation}
The rescaled process $(Z_{t} :t\ge 0)$ lies on the fixed domain $[0,L(0))$, such that \be Z_{t} \stackrel{dist}{=} B^{H}_{h(t)} \quad \mbox{for all} \quad t\ge0. \label{ZBfbm} \ee Applying the It{\^o} isometry (see \cite{Oksenda2003} page 29) on (\ref{sdefbm2}) and using (\ref{ZBfbm}) we have
\begin{equation} \langle (B^{H}_{h(t)})^{2} \rangle=\int_{0}^{t} \Big(K_{H}* \frac{L(0)}{L(\cdot)}(s')\Big)^{2} ds'  \nonumber \end{equation} and further using the isometry of $K_{H}$ (see \cite{Samko1993} page 187) gives \begin{equation} \langle (B^{H}_{h(t)})^{2}\rangle=H(2H-1)\int_{0}^{t}\int_{0}^{t} \frac{L(0)^{2}}{L(s)L(s')} |s-s'|^{2H-2} \, dsds' .\nonumber \end{equation} Since the process $\{B^{H}_{h(t)}\}_{t\ge 0}$ is a standard fBm with $\langle (B^{H}_{h})^{2} \rangle=h^{2H}$, we therefore have the following representation of the mapping on the domain $[0,L(t))$
\begin{equation} h_H (t)=\Big(H(2H-1)\int_{0}^{t}\int_{0}^{t} \frac{L(0)^{2}}{L(s)L(s')} |s-s'|^{2H-2} \, dsds'\Big)^{1/2H}.  \nonumber \end{equation}

\section*{References}



\end{document}